\documentclass[twocolumn,twocolappendix]{aastex631}

\usepackage{ascmac}
\usepackage{amsmath}
\usepackage{amssymb}
\usepackage{amsfonts}
\usepackage{bm}
\usepackage[T1]{fontenc}

\begin{document}

\title{Temporal Variation of Flare Occurrence Rates via the Spot Evolution on the Sun and Solar-type Stars}

\author[0000-0003-1579-5937]{Takato Tokuno}
\affiliation{ Department of Astronomy, School of Science, The University of Tokyo, 7-3-1 Hongo, Bunkyo-ku, Tokyo 113-0033, Japan }
\affiliation{ School of Arts \& Sciences, The University of Tokyo, 3-8-1 Komaba, Meguro, Tokyo 153-8902, Japan }
\email{tokuno-takato@g.ecc.u-tokyo.ac.jp}

\author[0000-0002-1297-9485]{Kosuke Namekata}
\affiliation{NASA Goddard Space Flight Center, 8800 Greenbelt Road, Greenbelt, MD 20771, USA}
\affiliation{The Catholic University of America, 620 Michigan Avenue, N.E. Washington, DC 20064, USA}
\affiliation{The Hakubi Center for Advanced Research/Graduate School of Science, Kyoto University, Kitashirakawa-Oiwake-cho, Sakyo-ku, Kyoto, 606-8502, Japan}
\affiliation{Department of Physics, Kyoto University, Kitashirakawa-Oiwake-cho, Sakyo-ku, Kyoto, 606-8502, Japan}
\affiliation{Division of Science, National Astronomical Observatory of Japan, NINS, Osawa, Mitaka, Tokyo, 181-8588, Japan}

\author[0000-0003-0332-0811]{Hiroyuki Maehara}
\affiliation{Okayama Branch Office, Subaru Telescope, National Astronomical Observatory of Japan, NINS, Kamogata, Asakuchi, Okayama
719-0232, Japan}

\author[0000-0002-1276-2403]{Shin Toriumi}
\affiliation{Japan Aerospace Exploration Agency, Institute of Space and Astronautical Science, 3-1-1 Yoshinodai, Chuo-ku, Sagamihara, Kanagawa 252-5210, Japan
3 Chiba University, Inage-ku, Chiba, Chiba 263-8522, Japan}

\begin{abstract}
The spot evolution on the Sun and solar-type stars is important for understanding the nature of consequential flaring activity. This study statistically investigates the variance of flare occurrence rate through the time evolution of spots on the Sun and solar-type stars. We have compiled the 28-year catalogs of solar flares and their source sunspots obtained from solar surface observations by NOAA and GOES for the Sun. Also, we combined the cataloged stellar flares with the time evolution of starspots estimated by light curves obtained by the 4-year \textit{Kepler} mission for solar-type stars. For the obtained 24124 solar flares and 180 stellar flares, we calculate the flare occurrence distribution with respect to $t_\mathrm{flare}-t_\mathrm{max}$, which represents the timing of flare through the spot evolution, where $t_\mathrm{flare}$ is the flare occurrence time, and $t_\mathrm{max}$ is the time when the source spot takes its maximum area. When normalized by the spot lifetime, we found that the flare occurrence distribution for $t_\mathrm{flare}-t_\mathrm{max}$ shows a similar distribution regardless of spot size or flare energy, suggesting that the Sun and the solar-type star share the same physical process in the spot-to-flare activity. On this basis, we propose a formula for the time variation of the flare occurrence rate per spot. Also, the correlation between the temporal variation of flare occurrence rate and the time evolution of spot area and the lack of difference in flare occurrence rate between the emergence and decaying phases provide a milestone for the nature of flare-productive spots.
\end{abstract}

\keywords{Solar flares(1496); starspots(1572); Stellar flares (1603); Sunspots (1653);  Flare stars (540); Solar analogs (1941)}


\section{Introduction} \label{sec:intro}

The magnetic activity of the solar surface has been a subject of active debate for a long time. One attractive target is sunspot and active region (AR). Long-term observations of sunspots are utilized to monitor the global process of solar activity \citep[e.g.,][]{Hathaway2015LRSP}. The formation and evolution of sunspots are governed by two mechanisms: emergence and decaying \citep[e.g.,][]{vanDriel-Gesztelyi2015LRSP}. While the emergence corresponds to the transport and amplification of magnetic flux generated by the dynamo mechanism in the solar interior via buoyancy and convection \citep[][]{Parker1955ApJ, Cheung2014LRSP}, the decay is caused by the diffusion of magnetic flux due to turbulence, near-surface flows, or reconnection \citep{Meyer1974MNRAS, Petrovay1997SoPh, Schrijver1998Natur, Kubo2008ApJ, RempelCheung2014ApJ}.

Another target is solar flare. They are explosive phenomena in solar atmospheres, thought to be caused by the conversion of magnetic energy into kinetic/thermal energy through magnetic reconnection \citep[e.g.,][]{Shibata2011LRSP}. The large flares are frequently accompanied by coronal mass ejections (CMEs), which eject massive amounts of coronal material into interplanetary space \citep[e.g.][]{Chen2011LRSP, Webb2012LRSP}. Since X-ray/EUV emissions from flares and CMEs are thought to impact planetary environments, understanding their properties is highly important from a planetary science \citep{Airapetian2016NatGe, Airapetian2020IJAsB, Yamashiki2019ApJ, Temmer2021LRSP}. 

By their nature, these two activities, spots and flares, are closely related phenomena. It is considered that the reconnection events causing flares occur near spots where magnetic flux is concentrated \citep[e.g.,][]{Priest2002AARv,Toriumi2019LRSP}. Indeed, statistical analyses reveal that strong solar flares emanate from large and complex sunspots \citep{Kunzel1960AN, Bornmann1994SoPh,Sammis2000ApJ, Lee2016ApJ, Toriumi2017ApJ}. Such studies to investigate the characteristics of flare-productive sunspots has attracted much attention as essential to understanding the spot-to-flare activity of the Sun.

Nowadays, the scope is not limited to the Sun. Thanks to recent space missions such as \textit{Kepler} \citep{Borucki2010Sci} and TESS \citep{Ricker2014SPIE}, we can access similar but surprisingly higher magnetic activities in solar-type stars (G-type main-sequence stars): gigantic starspots \citep{Notsu2013ApJ, Notsu2015PASJ, Reinhold2020Sci} and intensive flares called superflares \citep{Maehara2012Natur, Okamoto2021ApJ}. Despite their differences in magnitude, it is revealed that scaling laws with respect to the magnetic activities of the Sun are scalable to those of these stars both for flares \citep{Shibayama2013ApJS, Maehara2015EPS, Namekata2017ApJ, Tu2020ApJ, Tu2021ApJS, Vasilyev2024Sci} and spots \citep{Maehara2017PASJ, Namekata2019ApJ, Namekata2020ApJ}. Such comparative studies of the Sun and solar-type stars provide an opportunity to investigate a scale-independent framework to investigate the solar/stellar magnetic activities. Specifically, investigating the high magnetic activity of a slowly rotating solar-type star can lead to understanding and predicting extreme events that have occurred or will occur on the Sun \citep[e.g.,][]{Miyake2012Natur,Cliver2022LRSP,Usoskin2023SSRv, Obase2025ApJ}.

In light of the above, it is worth examining whether the relationship between spots and flares is common between the Sun and solar-type stars. It is known that stars with large starspots have a higher flare frequency \citep[e.g.][]{Okamoto2021ApJ,Tu2021ApJS}, but there is little evidence that giant starspots are the origin of stellar superflares due to the difficulty of spatially resolving the stellar disk. For example, it is known that there is a low correlation between the visibility of starspots and the frequency of flares \citep{Hawley2014ApJ,RoettenbacherVida2018ApJ,Doyle2018MNRAS,Doyle2020MNRAS, Feinstein2020AJ}. 
Conversely, \citet{AraujoValio2023MNRAS} have reported the correlation between them with detailed spot mapping, pointing to the importance of analysis incorporating time-directional information for linking starspot and stellar flare activity.

Despite this importance, most previous studies analyze snapshot information for starspots and have rarely considered time-directional information such as its evolution. From studies of the Sun, it is expected that superflares tend to occur when the magnetic flux is near the surface, there are few verification of superflare occurrences that can be linked to the time evolution of a starspot. In addition, few comparable solar analyses with solar-type stars that investigate the relationship between sunspot evolution and solar flare occurrence rate.

The main goal of this paper is to investigate the relationship between spot and flare for the Sun and solar-type stars in a comparable manner, taking into account the time-directional information. In this regard, we focus on the timing of flare occurrence relative to the time evolution of flare-productive spots.  A statistical study of this quantity allows us to verify the temporal variation of the flare rate with respect to the time evolution of a sunspot/starspot. Few previous studies have focused on this property of the Sun or solar-type stars, and our study is the first attempt.

The rest of the paper is organized as follows. In Section \ref{sec:method}, we describe the data we utilize and the method. Then, in Section \ref{sec:result}, we show several results of our statistical analysis. Finally, we discuss the results in Section \ref{sec:discussion}.

\section{Data and Analysis} \label{sec:method}

\begin{figure}[t!]
\centering
\plotone{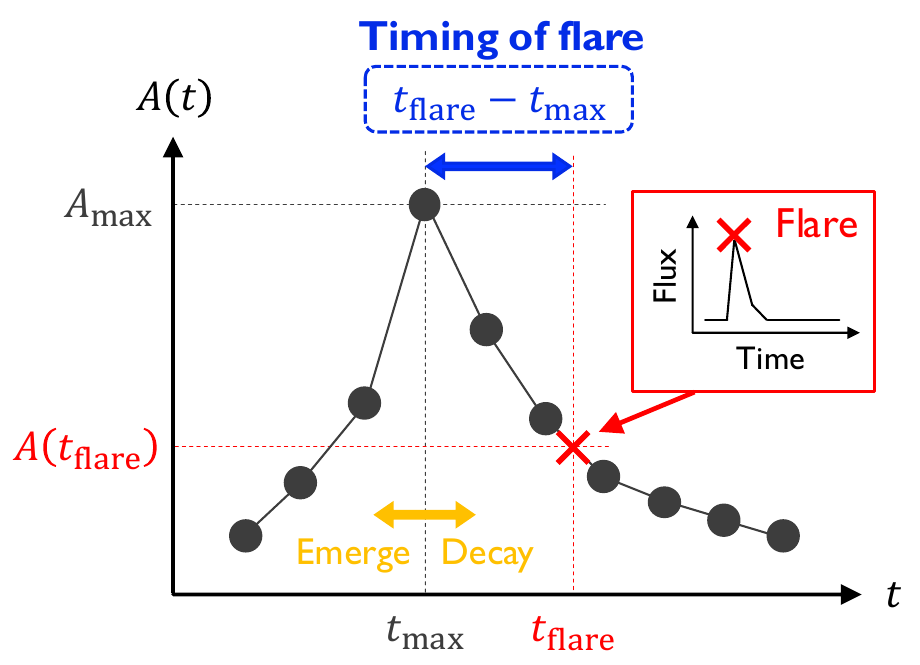}
\caption{Schematic picture to explain the parameters we deal with. The black circles with lines are the spot area (ordinate) at each time (abscissa) to show the time evolution of spot area, $A(t)$. The black dotted lines correspond to the time when the spot area takes the maximum, $t_\mathrm{max}$, and the maximum spot area, $A_\mathrm{max}\equiv A(t_\mathrm{max})$. The red symbol and red dotted lines indicate the time when the flare occurred (see inset), $t_\mathrm{flare}$ and the spot area at that time, $A(t_\mathrm{flare})$. The parameter highlighted by blue color at the top, $t_\mathrm{flare}-t_\mathrm{max}$, means the timing of flare during spot evolution, which is the parameter we focus on. The yellow arrow and characters explain that the time satisfying $t-t_\mathrm{max}<0$ and $t-t_\mathrm{max}>0$ represent the emergence and decaying phases, respectively.
\label{fig:schematic}}
\end{figure}

This section shows the analysis to determine when flares occur during spot evolution on the Sun and the solar-type stars. 

Here, we introduce the parameters we utilize, which are schematically summarized in Figure \ref{fig:schematic}. We define the occurrence time, $t_\mathrm{flare}$, and its energy, $E_\mathrm{flare}$, as the observed quantities of flare. We also define the evolution of spot area $A(t)$ as the function of time $t$ is acquired for their source spots. When defining $A_\mathrm{max} \equiv A(t_\mathrm{max})$ and $t_\mathrm{max}$ as the maximum value of $A(t)$ and the time taking it, respectively, we can regard the time $t$ with $t-t_\mathrm{max} > 0$ ($t-t_\mathrm{max} < 0$) as the emergence (decaying) phase. On this basis, we consider that  $t_\mathrm{flare}-t_\mathrm{max}$ represents the timing of flares through the spot evolution. Our aim is to discuss the frequency of flare occurrences with respect to $t_\mathrm{flare}-t_\mathrm{max}$, combining other flare and spot features: the flare energy $E_\mathrm{flare}$, the maximum spot area $A_\mathrm{max}$, and the spot area at the flare occurrence time $A(t_\mathrm{flare})$. 

We conducted two types of analyses to obtain these parameters for each flare and its source spot. The one is a solar analysis using the archive of sunspots and flares based on spatially resolved solar disk observations (Section \ref{subsec:method_solar}). The other is an analysis using the light curves for solar-type stars observed by the \textit{Kepler} mission. We explain how to compile flare-productive starspot candidates based on the \textit{Kepler} light curve in this analysis in Section \ref{subsec:method_stellar}. Finally, in Section \ref{subsec:method_normalize}, we describe the normalization procedure to compare flares with different energies and spots with different sizes. 

We note the symbols and units we utilize hereafter. The symbol $\odot$ is a subscript indicating the solar value. As the unit of spot area, we use the millions of a solar hemisphere (MSH; $1 \, \mathrm{MSH} = 10^{-6} \times 2\pi R_\odot^2 = 3.1 \times  10^{16} \, \mathrm{cm}^2$), where $R_\odot$ is the solar radius. 

\subsection{Solar Analysis}
\label{subsec:method_solar}

\begin{table}[t!]
    \centering
    \caption{Number of flares ($n_\mathrm{flare}$), number of flares with identified source spots ($n_\mathrm{flare+spot}$), and the detection rate of source spots ($n_\mathrm{flare+spot}/n_\mathrm{flare}$) obtained by our analysis: the 28-year NOAA data (Section \ref{subsec:method_solar}) and 4-year \textit{Kepler} data (Section \ref{subsec:method_stellar}). }
    \label{tab:number_of_flares}
    \begin{tabular}{cccc} 
	\hline
      &  $n_\mathrm{flare}$ &  $n_\mathrm{flare+spot}$ & $n_\mathrm{flare+spot}/n_\mathrm{flare}$ \\
    \hline
    \textbf{Solar Flares} & \textbf{32378} & \textbf{24124} & \textbf{75} \% \\
    \hline
    C1-C3 & 17698 & 12252 & 69 \% \\
    C3-M1 & 10973 & 8607 & 78 \% \\
    M1-M3 & 2661 & 2302 & 87 \%  \\
    M3-X1 & 801 & 720 & 90 \%  \\
    >X1 & 245 & 243 & 99 \%  \\
    \hline
    \textbf{Stellar Flares} & \textbf{308} & \textbf{180} & \textbf{58} \% \\
    \hline
    \end{tabular} 
\end{table}

We compiled the flare and sunspot data distributed by the National Oceanic and Atmospheric Administration (NOAA). NOAA identifies each sunspot group by using the NOAA Active Region (AR) number. Hereafter, we conducted analyses using each sunspot group defined by its NOAA AR number as the fundamental unit of a sunspot.

For flares, we utilize the 28-year solar flare data observed by the Geostationary Operational Environmental Satellite (GOES) mission from January 1996 to September 2024. For the flare after August 1996, we utilized the daily reports named the Event Report\footnote{\url{https://www.ngdc.noaa.gov/stp/space-weather/swpc-products/daily_reports/solar_event_reports/}}, and for the flare before August 1996, we utilized the annual reports named the XRS Reports\footnote{\url{https://www.ngdc.noaa.gov/stp/space-weather/solar-data/solar-features/solar-flares/x-rays/goes/xrs/}}. From these data, we extracted the occurrence time and date, GOES X-ray class, and NOAA AR number that has been identified as the flare source for each flare. We note that the GOES X-ray class is the classification of solar flares according to the peak X-ray (1-8 \AA) flux ($F_\mathrm{X}$) measured by the GOES mission\footnote{$F_\mathrm{X}$ of A, B, C, M, and X class solar flares are $< 10^{-7}$, $10^{-7}$-$10^{-6}$, $10^{-6}$-$10^{-5}$, $10^{-5}$-$10^{-4}$ and $\geq 10^{-4} \, \mathrm{W/m^2}$, respectively. Also, each X-ray class is divided into a linear scale from 1 to 9.}. In this analysis, we analyzed flares equal to or greater than the C1 class, which is not considered to be masked by background flux. When converting $F_X$ into $E_\mathrm{flare}$, we use the empirical relation $E_\mathrm{flare} = 10^{30} (F_\mathrm{X}/10^{-5} \, \mathrm{W/m^2}) \, \mathrm{erg}$, which is suggested by the solar observation \citep[e.g.,][]{Emslie2012ApJ}. 

For sunspots, we use the 28-year sunspot data observed by the United States Air Force from January 1996 to September 2024, which are reported by the daily reports named the Solar Region Summary. We extracted the NOAA AR number, hemispheric location, and spot size for each observation date from these reports. By performing linear interpolation, we finally obtain $A(t)$ for each sunspot group\footnote{We note that it tends to the underestimate the sunspot area near the limb of the solar disk because this sunspot data is analyzed manually (see also Sections \ref{subsec:result_frequency} and \ref{subsec:result_property}).}. Here, we exclude sunspots with less than 30 MSH and sunspots with less than two data points. Also, we assume that the observation times is all at noon. 

It should be mentioned that the NOAA AR number does not take into account recurrent spots, which are sunspots that turn into the backside but appear on the front side again. That is to say, all sunspots that appear on the front side are newly numbered, even if they are identical to an already-numbered sunspot. To account for the recurrent spots, we treat NOAA AR numbers that meet certain criteria as representing the same sunspot group. The utilized criteria were that their location is similar and that their evolution does not deviate from expected spot evolution. Their details are provided in Appendix \ref{app:recurrent_sunspot}. 

As a result, we obtained $t_\mathrm{flare}-t_\mathrm{max}$, $A(t_\mathrm{flare})$, $t_\mathrm{max}$, and $A_\mathrm{max}$ for 24124 solar flares and their source sunspot. Table \ref{tab:number_of_flares} shows the number of flares $n_\mathrm{flare}$, the number of flares whose source spot is identified $n_\mathrm{flare+spot}$, and their ratio for each X-ray class. 

\subsection{Stellar Analysis} \label{subsec:method_stellar}

\begin{figure*}[t!]
\centering
\plotone{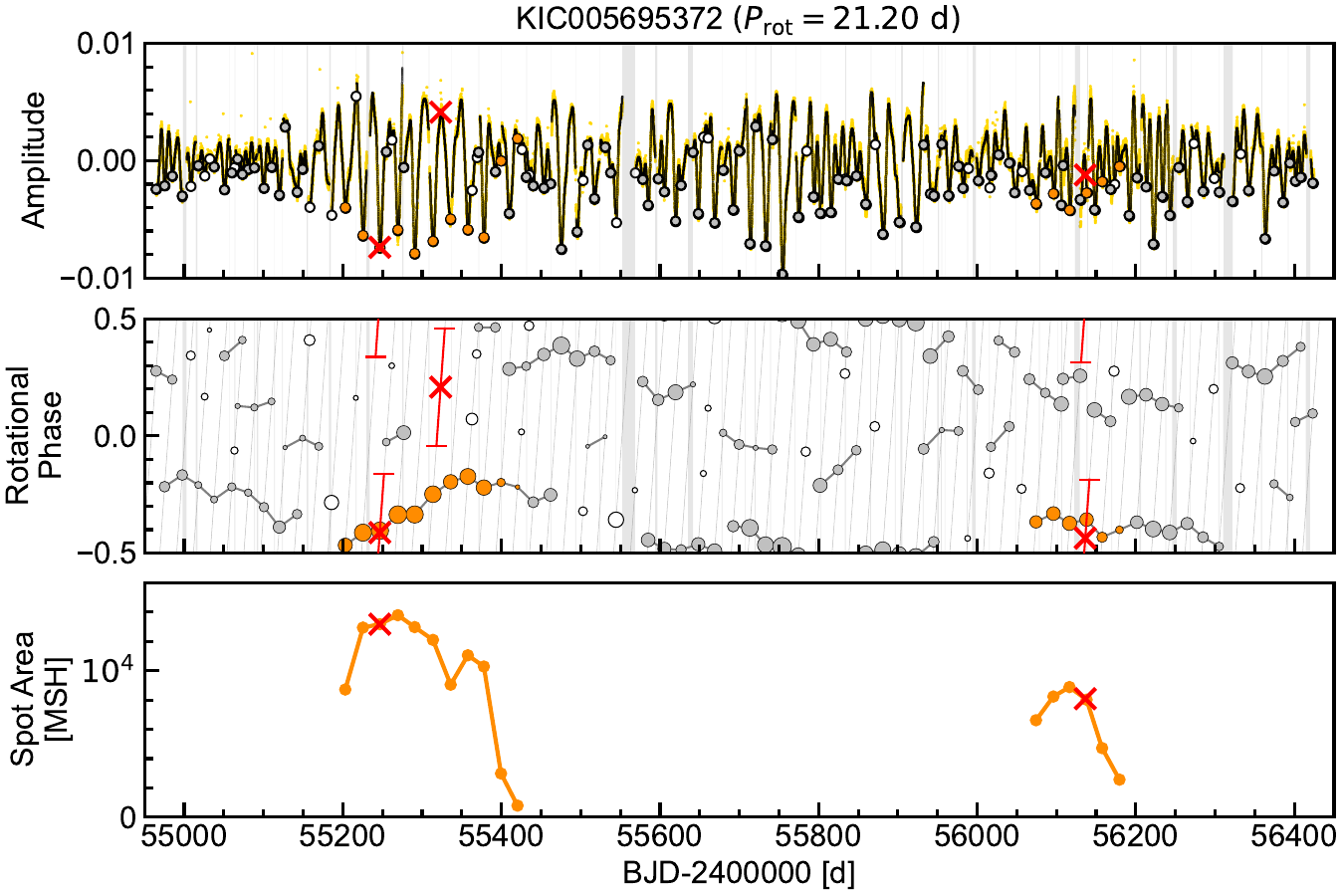}
\caption{Successfully analyzed examples of the time evolution of flare-productive starspot candidates and the timing of flare. We note that we adopt $\varepsilon=0.10$ and $P_\mathrm{rot}=21.20 \, \mathrm{d}$ for the clustering analysis of this star, KIC 5695732.
(Upper panel) \textit{Kepler} light curve for Barycentric Julian Date (BJD) subtracting 2450000 d. The background yellow line is the observed \textit{Kepler} light curve, and the black line is the smoothed one. The vertical gray-shaded regions correspond to the observational gaps longer than 5 hr. As for symbols and colors, the same one applies to the following panels: The open circles represent the unclustered local minima of the light curve; the orange-filled circle is the clustered local minima corresponding to the flare-productive starspot candidates; the gray-filled circle is the clustered local minima corresponding to other spots; and the red symbol indicates positions where flares were detected on the light curve.
(Middle panel) Phase-time diagram, where the vertical axis corresponds to phases of the local minima detected in the upper panels compared to the rotational period. The size of the circles represents the depth of each minimum from the nearby local maxima. The gray shaded line represents the direction in which time is progressing. In this figure, a red line has been added to each red symbol representing flare to indicate the criteria that pick up the candidates for flare sources. 
(Bottom Panel) Time evolution of spot area estimated by the depth of the local minima for the spots we focus on. We also show when the flares occurred in this evolution.
\label{fig:stellar_sample}}
\end{figure*}

In the stellar analysis, we utilized the data of flares and starspots of solar-type stars obtained by the \textit{Kepler} light curves. After introducing the sample selection of stars and flares, we explain the analysis to estimate the properties of spots as the source of flares from the \textit{Kepler} light curves. We show an example of a successful analysis in Figure \ref{fig:stellar_sample}, which helps the reader to comprehend the analysis visually.

The stellar and flare data in this study follow those of \citet{Okamoto2021ApJ}. The sample of \citet{Okamoto2021ApJ} consists of the stars whose effective temperature of the photosphere ($T_\mathrm{phot}$), stellar radius ($R_\mathrm{star}$), and evolutionary stage  are estimated by \citet{Berger2018ApJ}\footnote{They are estimated by the stellar parameter estimated by \citet{Pinsonneault2012ApJS} and \citet{Mathur2017ApJS} using \textit{Kepler} Data Release 25 \citep{ Thompson2016KSCI} and by the parallax of \textit{Gaia} Data Release 2 \citep{GaiaCollab2018AA}.} and whose rotation periods ($P_\mathrm{rot}$) are estimated by \citet{McQuillan2014ApJS}\footnote{They are estimated by the quasi-periodic brightness variation using the auto-mated autocorrelation-based method.}, and that satisfies the main-sequence phase and $T_\mathrm{phot}=5100 \textrm{-} 6000 \, \mathrm{K}$.  
\citet{Okamoto2021ApJ} conducted a comprehensive search of \textit{Kepler} 30-minute time-cadence data for those stars and compiled the stellar flares from those stars, which are observed as intensive peaks on the light curves. From the catalog of \citet{Okamoto2021ApJ}, we extract the Kepler Input Catalog (KIC) number, $T_\mathrm{phot}$, $R_\mathrm{star}$, $P_\mathrm{rot}$ for each flare-source solar-type star and $t_\mathrm{flare}$, $E_\mathrm{flare}$ for each flare. 

Among these stars, we specifically focus on the slowly rotating solar-type stars, which would have magnetic activity similar to the solar one. The reason for this extraction is that it allows for a more direct comparison between the Sun and solar-type stars (see Section \ref{sec:intro}) and also because the evolution of long-lived starspots is difficult to track in subsequent analyses. As a result, we extracted 308 flares from 140 solar-type stars filling $P_\mathrm{rot} \geq 10 \, \mathrm{d}$.

Hereafter, we introduce the method to estimate the property of the flare-productive starspot candidates. We conduct the local-minima tracing method, which is a similar analysis to that of \citet{Hall1994IAPPP}, \citet{Henry1995ApJS}, \citet{Davenport2015PhDT}, and \citet{Namekata2019ApJ, Namekata2020ApJ}. The light curve of a rotating star with starspots shows the quasi-periodic variation, which is due to spots appearing into and out of view by rotation. The local minima of the light curve can be regarded as a situation where the starspot is located on the meridian of the stellar disk \citep{Notsu2013ApJ,Namekata2019ApJ, Namekata2020ApJ}. Thus, we consider each starspot can be identified and parameterized by tracing the rotational phases and depths of local minima. When local minima are observed over multiple rotational cycles, we can track the time evolution of these recurrent starspots. On the basis of this method, we extract the flare-productive starspot candidates, detecting local minima appearing close to the flare occurrence time. The specific procedures are as follows. 

In the beginning, we detect local minima from the \textit{Kepler} light curve. By cutting off high-frequency components of the light curve via the discrete Fourier transformation, we obtained the smoothed light curve from observed one (see also upper panel of Figure \ref{fig:stellar_sample}). Here, the cut-off frequency was set to $4/P_\mathrm{rot}$ to avoid over- and under-smoothing. After that, we detected the local minima of the light curve using the \texttt{signal.argrelmin} in the Python package \texttt{Scipy} \citep{Virtanen2020NatMe}. The number of points on each side used to detect local minima was set to 20. As a result, we can obtain two features of local minimum: its appearance time ($t_\mathrm{lm}$) and depths of normalized flux ($\Delta F/F$). We treat the detection times of local minima as $t_\mathrm{lm}$. Also, we utilize the local depth of the local minima from the nearby local maxima as $\Delta F/F$, following \citet{Maehara2017PASJ}. 

Next, we estimate the starspot coverage from  the depths of local minima. Following \citet{Maehara2017PASJ}, given the effective temperature of spot ($T_\mathrm{spot}$), we can calculate the spot area $A(t_\mathrm{lm})$ as 
\begin{align}
    A(t_\mathrm{lm}) = \left( \dfrac{\Delta F}{F}\right) \left( \dfrac{R_\mathrm{star}}{R_\odot} \right)^2  \frac{T_\mathrm{phot}^4}{T_\mathrm{phot}^4-T_\mathrm{spot}^4}. \label{eq:A-Tspot}
\end{align}
Also, \citet{Maehara2017PASJ} derived the empirical relation between $T_\mathrm{spot}$ and  $T_\mathrm{phot}$ as
\begin{align}
\frac{T_\mathrm{spot}}{10^3 \, \mathrm{K}} = - 35.8 \left( \frac{T_\mathrm{phot}}{10^3 \, \mathrm{K}}\right)^2 +751 \left( \frac{T_\mathrm{phot}}{10^3 \, \mathrm{K}} \right)+ 808, \label{eq:Tspot-Tphot}
\end{align}
which is based on the result of the Doppler imaging technique \citep{Berdyugina2005LRSP}. Thus, by substituting $T_\mathrm{phot}$ and $R_\mathrm{star}$ into Equations (\ref{eq:A-Tspot}) and (\ref{eq:Tspot-Tphot}), we obtained $A(t_\mathrm{lm})$ for each local minimum. Although this approach cannot distinguish a gigantic spot or a spot group of large starspots in nearby longitude, \citet{Namekata2019ApJ} and \citet{Namekata2020ApJ} validated by using solar and stellar data that the local minima method tends to trace the largest sunspot/starspot in the hemisphere appearing at the time (see also Section \ref{subsec:discussion_uncertainty}).

When we define rotational cycles ($m \in \mathbb{Z}$) and phases ($\ell \in \mathbb{R}$) satisfying $t_\mathrm{lm} = (m + \ell) P_\mathrm{rot}$ and $-0.5 \leq \ell < 0.5$ for each local minimum, it is considered that the local minima corresponding to long-lived recurrent spots have the similar phase over multiple cycles. Thus, we perform clustering analysis for rotational cycles and phases to identify these recurrent spots. For clustering, we use \texttt{DBSCAN} in the Python package \texttt{scikit-learn} \citep{Pedregosa2011JMLR}. This commonly utilized clustering algorithm, in brief, finds the core points that have more than $n_\mathrm{cluster}$ satellite points within the length of $\varepsilon$ and detects clusters by connecting the core points with each other. This package allows for fast analysis and is useful for statistical discussions. In this analysis, we treated the local minima clustered in the two-dimensional space $(m, C \ell)$ by \texttt{DBSCAN} with $n_\mathrm{cluster}=2$ as the candidate of recurrent starspots, where $C$ is a constant introduced to control the clustering area in the space. The criteria for identifying satellite points in \texttt{DBSCAN} are controlled by two parameters, $\varepsilon$ and $C$. In this study, we set them up so that local minima within 0.1 of the rotation phase in adjacent rotational cycles are recognized as satellite points, i.e., clustered. We note that we treated clusters with multiple peaks in the curve connecting $A(t_\mathrm{lm})$ as two different clusters each as a separated starspot origin. The detailed procedures to determine control parameters and to decrease false positives are described in Appendix \ref{app:Clustering}.

Finally, we explain the method to distinguish whether a starspot is a flare source or not, using the difference between $t_\mathrm{flare}$ and $t_\mathrm{lm}$. Given that $t_\mathrm{lm}$ represent the time when the starspot is located in the meridian of the stellar disk, we can consider that the time in $[t_\mathrm{lm}-P_\mathrm{rot}/4, t_\mathrm{lm}+P_\mathrm{rot}/4]$ can be interpreted as the period when a starspot is visible on the stellar disk. If a flare occurs during this period, the starspot corresponding to the local minima is considered to be a candidate for the source of the flare. On the basis of the above discussion, we label a starspot where there is only one local minimum that satisfies $|t_\mathrm{lm}-t_\mathrm{flare}| < P_\mathrm{rot}/4$ as the flare-productive starspot candidates (see also middle panel of Figure \ref{fig:stellar_sample}). 

Combining these results, we can obtain the time evolution of flare-productive starspot candidates and the timing of flare occurrences. For the clustered local minima, $A(t)$ was obtained by linear interpolation of multiple $A(t_\mathrm{lm})$, and, thereby, $A_\mathrm{max}$ and $t_\mathrm{max}$ were obtained. For the unclustered local minima, $A_\mathrm{max}$ and $t_\mathrm{max}$ were obtained, assuming that the observed point represents the maximum spot area. We then calculated $t_\mathrm{flare}-t_\mathrm{max}$ and $A(t_\mathrm{flare})$ using them. (see also bottom panel of Figure \ref{fig:stellar_sample}) 

As a result, we obtained $t_\mathrm{flare}-t_\mathrm{max}$, $A(t_\mathrm{flare})$, $t_\mathrm{max}$, and $A_\mathrm{max}$ for 180 stellar flares on 89 solar-type stars (see also Table \ref{tab:number_of_flares}).

\subsection{Normalization} \label{subsec:method_normalize}

\begin{figure}[t!]
\centering
\plotone{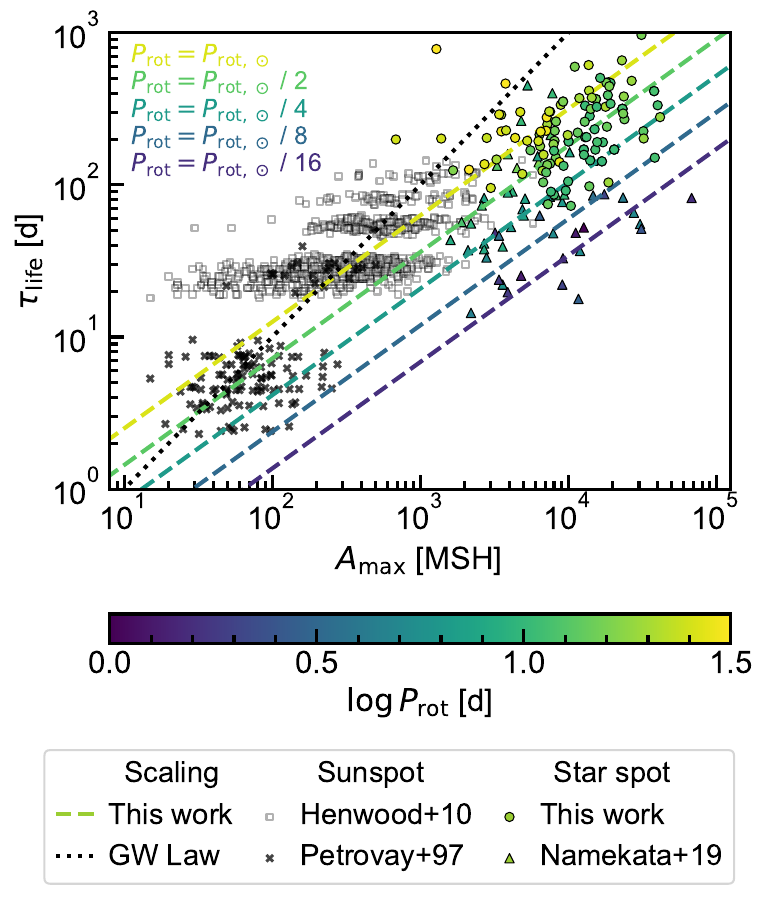}
\caption{Comparison between maximum spot area and lifetime of sunspots and starspots with scaling laws. The colored circles and triangles represent the value of the starspots used in our study and \citet{Namekata2019ApJ}, respectively. The colors of symbols correspond to $P_\mathrm{rot}$, as shown in the color bar. Also, the gray squares and the black crosses are the values of sunspots in \citet{Petrovay1997SoPh} and \citet{Henwood2010SoPh}, respectively. 
The colored lines show the scaling law used in this work (Equation \ref{eq:tau_scaling}) for $P_\mathrm{rot}=P_\mathrm{rot,\odot}$, $P_\mathrm{rot,\odot}/2$, $P_\mathrm{rot,\odot}/4$, $P_\mathrm{rot,\odot}/8$, and $P_\mathrm{rot,\odot}/16$. The colors of the lines are consistent with the color bar. We also show the Gnevyshev–
Waldmeier law \citep[$\tau_\mathrm{life}/10\, \mathrm{d}=A_\mathrm{max}/1\, \mathrm{MSH}$;][]{Gnevyshev1938IzPul,Waldmeier1955} as the black dotted line for reference.
\label{fig:lifetime}}
\end{figure}

The spot data we obtained are distributed over a wide range, with the range of $A_\mathrm{max} \simeq 10^{0.5 \textrm{-}3.5}$ MSH for the sunspot and $A_\mathrm{max} \simeq 10^{2.5 \textrm{-} 4.5}$ MSH for starspots. Generally, larger spots have longer evolutionary timescale \citep[see, e.g., Figure 10 of ][and therein references]{Namekata2020ApJ}. In order to compare spots of different sizes, we conduct normalization with respect to the time. Assuming spot evolution is governed by similar physics and evolves analogously regardless of size, it is plausible to normalize $t_\mathrm{flare}-t_\mathrm{max}$ and $A(t_\mathrm{flare})$ by the spot lifetime ($\tau_\mathrm{life}$) and  $A_\mathrm{max}$, respectively. While we can do the former normalization from the obtained data, the latter is not. Even for sunspots, it is challenging to determine $\tau_\mathrm{life}$ accurately since there are many sunspots that arise and vanish on the backside. Measurements of the $\tau_\mathrm{life}$ of starspots are more difficult due to the sparseness of the observed points. 

On this basis, we normalize $t_\mathrm{flare}-t_\mathrm{max}$ using  estimated $\tau_\mathrm{life}$ from the scaling law. Considering the dependence on $P_\mathrm{rot}$ reported in \citet{Namekata2019ApJ},  we utilize the power-law relation between $\tau_\mathrm{life}$, $A_\mathrm{max}$, and $P_\mathrm{rot}$ as
\begin{align}
    \tau_\mathrm{life} = 0.5 \left( \frac{A_\mathrm{max}}{1 \, \mathrm{MSH} } \right)^{p} \left( \frac{P_\mathrm{rot}}{P_\mathrm{rot,\odot} } \right)^{q} \, \mathrm{d}, \label{eq:tau_scaling}
\end{align}
where $P_\mathrm{rot,\odot}$ is the solar rotation period and is set to $26 \, \mathrm{d}$ here. Also, $p$ and $q$ are the power-law indexes and we adopted $p=0.7$ and $q=0.8$ in this study. Figure \ref{fig:lifetime} shows the comparison of Equation (\ref{eq:tau_scaling}) with the value of sunspots in \citet{Petrovay1997SoPh} and \citet{Henwood2010SoPh} and with the value of starspots in \citet{Namekata2019ApJ} and this study. The method to estimate $\tau_\mathrm{life}$ in this study is based on the scaling law of spot evolution, where the detail is described in Appendix \ref{app:scaling_tau}. We can confirm that Equation (\ref{eq:tau_scaling}) is roughly consistent with the trend of these observations. 

We note that the power-law index of $A_\mathrm{max}$ in Equation (\ref{eq:tau_scaling}), $p=0.7$, is retrieved from the suggestion of \citet[][]{Namekata2019ApJ, Namekata2020ApJ} based on the starspot analysis. 
However, the physical basis of this value remains uncertain \citep[see also][]{Namekata2019ApJ}. In fact, various physical processes have been proposed to account for the decay process --- which typically dominates the sunspot lifetime --- including surface flows \citep{Kubo2008ApJ, RempelCheung2014ApJ, Rempel2015ApJ}, turbulent diffusion \citep{Meyer1974MNRAS, Krause1975SoPh}, turbulent erosion \citep{Petrovay1997SoPh}, shear flow caused by differential rotation \citep{Hall1994IAPPP, Henry1995ApJS,Isik2007A&A}, and supergranulation \citep{SimonLeighton1964ApJ, BradshawHartigan2014ApJ}. Depending on the relative importance of these processes, the value of $p$ may exhibit complex behavior.

On the other hand, we adopted the positive power-law index of $P_\mathrm{rot}$, $q=0.8$, as a value that also roughly explains the dispersion in Figure \ref{fig:lifetime}. The positive correlation between $A_\mathrm{max}$ and $P_\mathrm{rot}$ might be attributed to the fast surface flow from the strong differential rotation \citep{HottaYokoyama2011ApJ,Reinhold2015A&A,BalonaAbedigamba2016MNRAS,Brun2022ApJ} or high magnetic diffusivity \citep{BradshawHartigan2014ApJ} in fast-rotating stars. However, as in the above case, the theoretical validation is challenging due to the unclear nature of the sunspot decay.

In summary, it is fair to say that there is currently no clear support for specific values of $p$ and $q$. It suggests that relying on the empirical relation given by Equation (\ref{eq:tau_scaling}) represents the best available approach for this study and that further theoretical investigations are required for improvement.

\section{Result}
\label{sec:result}

\subsection{Scatter plot}
\label{subsec:result_scatter}

\begin{figure*}[t!]
\centering
\plotone{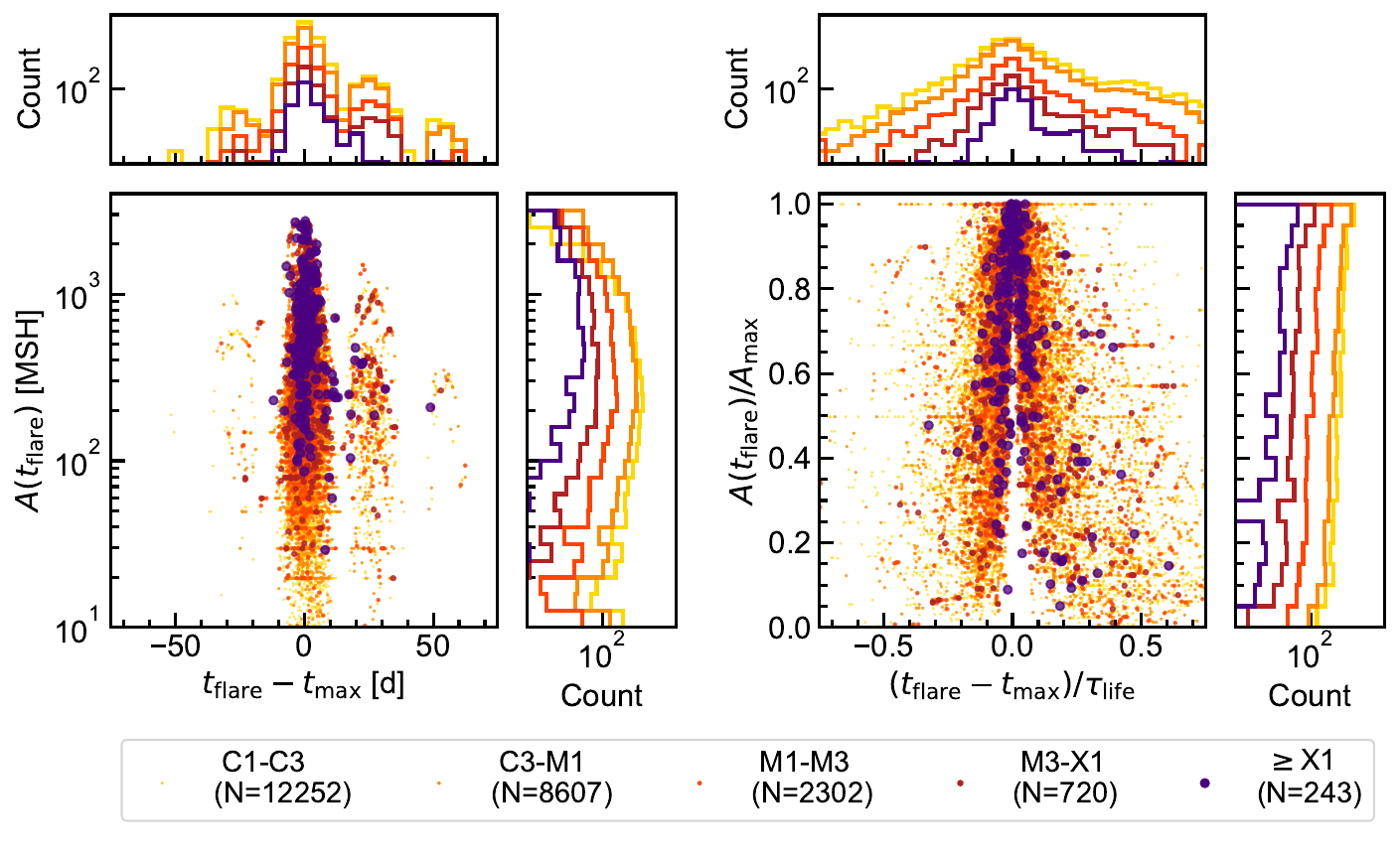}
\caption{Scatter plot of the occurrence time of solar flare $t_\mathrm{flare}-t_\mathrm{max}$ with sunspot area at that time $A(t_\mathrm{flare})$. While the left panel shows the observed value, the right panel shows the normalized value, where the spot area and the timing of flare are normalized by maximum spot size $A_\mathrm{max}$ and the spot lifetime $\tau_\mathrm{life}$, respectively (see Section \ref{subsec:method_normalize}). The color and size of symbols represents the GOES X-ray classes (C1-C3, C3-M1, M1-M3, and after X1 class, see legend). The sample size is shown in the legend. For each class, the histograms for each parameter are attached along the corresponding axis with matching colors. 
\label{fig:scatter_solar}}
\end{figure*}

\begin{figure*}[t!]
\centering
\plotone{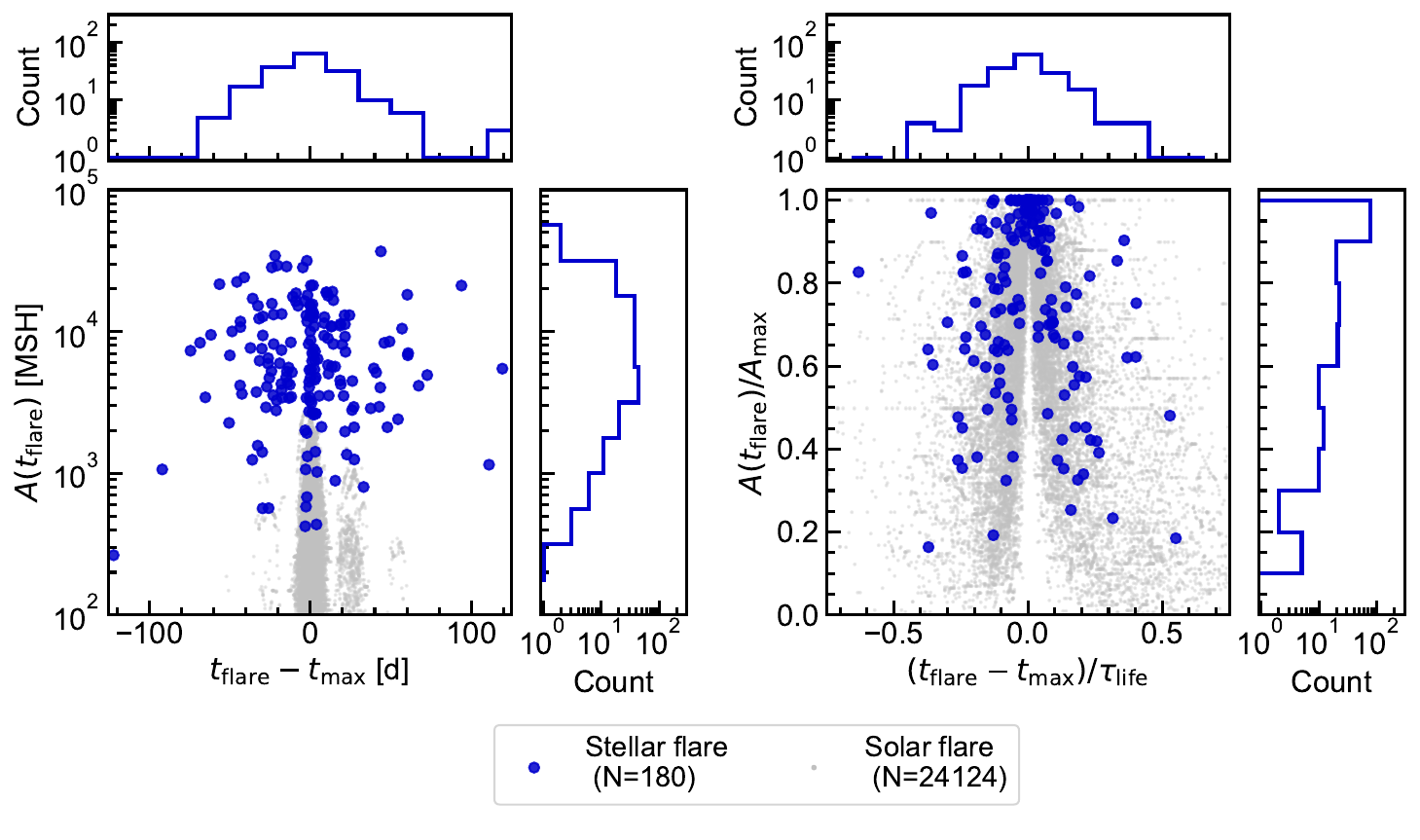}
\caption{The same figure as Figure \ref{fig:scatter_solar} but for the starspot and stellar flare. The blue symbols represent the value of stellar flare. The gray symbols represent the values of solar flares, which are the same as Figure \ref{fig:scatter_solar}.
\label{fig:scatter_stellar}}
\end{figure*}

Figure \ref{fig:scatter_solar} shows the scatter plots of $t_\mathrm{flare}-t_\mathrm{max}$ and $A(t_\mathrm{flare})$ for solar flares with different colors for each GOES X-ray class. The Left panel shows the observed values, and the right panel shows the normalized values (see Section \ref{subsec:method_normalize}). 
From two scatter plots, We found that while the observed solar values (left panel of Figure \ref{fig:scatter_solar}) show the energy-dependent behavior that flares with higher energy arise from larger sunspots, the solar normalized values (right panel of Figure \ref{fig:scatter_solar}) exhibit  similar scatter plots regardless of the energy due to the analogousness of spot evolution. The attached histograms support these trends. This confirms the validity of the normalization described in Section \ref{subsec:method_normalize} in the solar case. 

In the scatter plot and attached histogram in the left panel of Figure \ref{fig:scatter_solar}, a sparse area appears in $|t_\mathrm{flare}-t_\mathrm{max}| \simeq 10\textrm{-}20 \, \mathrm{d}$ and $|t_\mathrm{flare}-t_\mathrm{max}| \simeq 40\textrm{-}50 \, \mathrm{d}$. This corresponds to the period when sunspots turn to the backside (cf. $P_\mathrm{rot, \odot} \simeq 26 \, \mathrm{d}$). On the other hand, this tendency is weakened in the normalized case (right panel of Figure \ref{fig:scatter_solar}). This occurs because the time intervals corresponding to the sparse regions are distributed across different normalized times, which statistically dilutes this effect. Since the source spot of stronger flares is limited to larger ones (see the left panel of Figure \ref{fig:scatter_solar}), the dilution is weaker for larger flares. 

Figures \ref{fig:scatter_stellar} is a similar scatter plots to Figure \ref{fig:scatter_solar} but for stellar flares. From the observed value (left panel of Figure \ref{fig:scatter_stellar}), it can be seen that flare-productive starspot candidates of solar-type stars in our sample have larger size and longer lifetime than sunspots (see also Figure \ref{fig:lifetime}). On the other hand, we found that the scattering region of stellar normalized values overlaps with that of solar normalized values (right panel of Figure \ref{fig:scatter_stellar}). It indicates that our normalization is valid even when the stellar rotation, spot size, and spot lifetime are different. As a difference from the case of solar, the sparse area due to the recurrent spot does not appear in the stellar case (Figure \ref{fig:scatter_solar}). We consider this to be due to the duration when the starspots turn to the backside; that is, $P_\mathrm{rot}/2$ varies from star to star.

For both the Sun and solar-type stars, the histograms attached to the horizontal axes in the right panels of Figures \ref{fig:scatter_solar} and \ref{fig:scatter_stellar} suggest that most flares occur when spots are near their peak. Additionally, the shapes of these histograms for the horizontal axis appear to be almost energy-independent (see Section \ref{subsec:result_frequency} for more detail). We note that these trends are consistent with those of histograms attached to the vertical axis in the right panels of Figures \ref{fig:scatter_solar} and \ref{fig:scatter_stellar}, showing that the flares are likely to occur when $A(t_\mathrm{flare})/A_\mathrm{max} \simeq 1$ and that the shapes of the histograms are almost energy-independent.  

\subsection{Occurrence Frequency Distribution of Flare Timing}
\label{subsec:result_frequency}

\begin{figure*}[t!]
\centering
\plotone{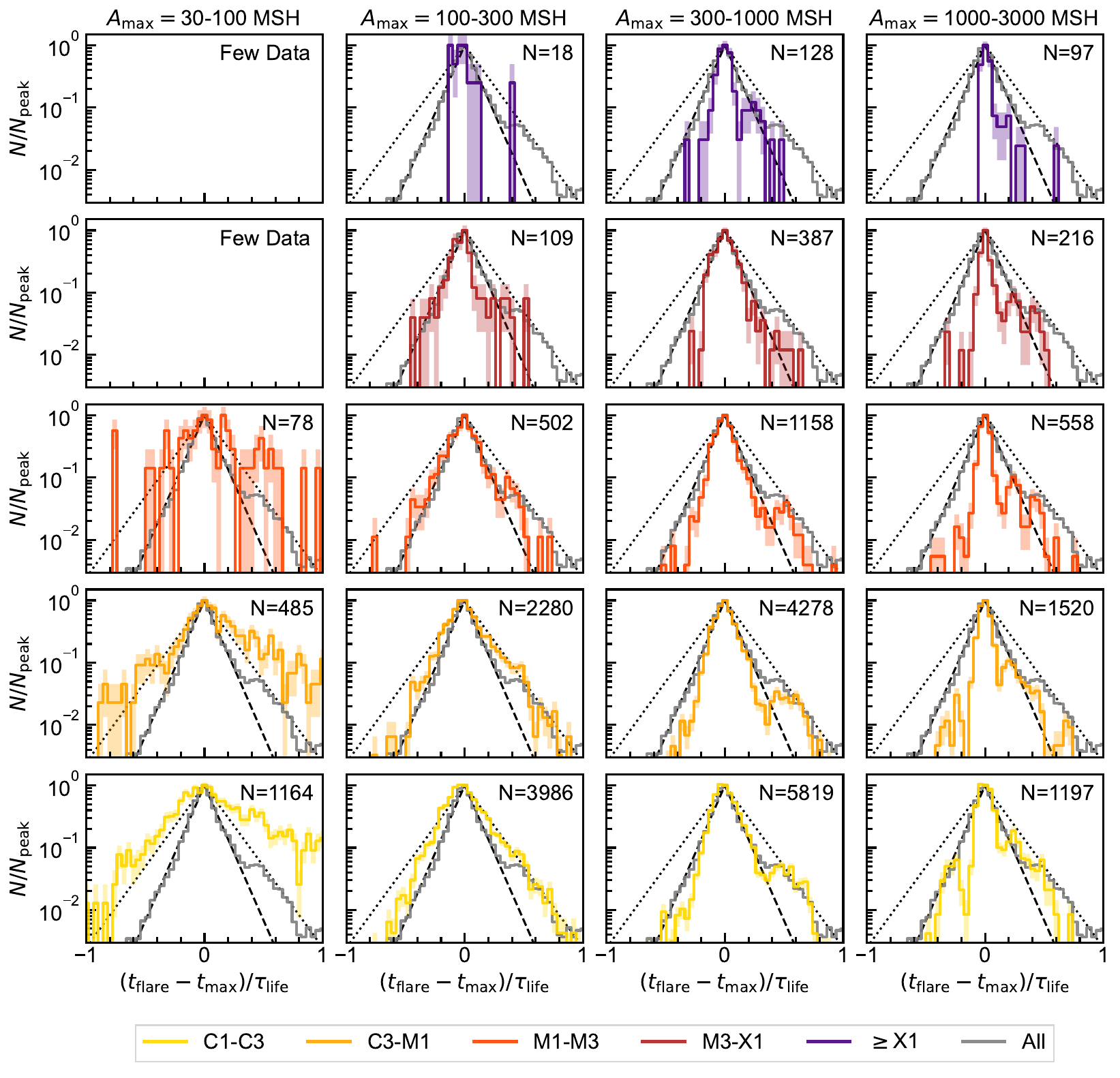}
\caption{Frequency distributions of occurrence time on the solar flares with different sunspot sizes and flare magnitudes. For all panels, we use the same axes: the abscissas represent the normalized flare timing $(t_\mathrm{flare}-t_\mathrm{max})/\tau_\mathrm{life}$ (see Section \ref{subsec:method_normalize}), and the ordinates represent the occurrence frequency of flares per year for each bin normalized by the value at the peak of spot evolution $N/N_\mathrm{peak}$. Each row, column, and color of panels represent the condition to divide sample: each row represents $A_\mathrm{max}$ of source spots ($30\textrm{-}100$, $100\textrm{-}300$, $300\textrm{-}1000$, and $1000\textrm{-}3000 \, \mathrm{MSH}$ from left to right, respectively; see also the topmost text); and each row and color represent the GOES X-ray classes (C1-C3, C3-M1, M1-M3, and after X1 class from top to bottom, respectively; see also the legend). The number of samples is shown in the upper right corner of each panel except when there are less than five samples (shown as 'Few Data'). The gray and colored lines in each panel, respectively, show the histogram for whole solar flares and for extracted solar flares. For the latter, its 1-$\sigma$ error due to sample size for each bin is described as the color-shaded region. The black dotted and dashed lines in each panel represent the fitting model $N/N_\mathrm{peak}$ written as Equation (\ref{eq:reference_freq}) setting $\alpha_\mathrm{e,d}=6$ and $\alpha_\mathrm{e,d}=10$, respectively.
\label{fig:flare_timing_solar}}
\end{figure*}

\begin{figure*}[t!]
\centering
\plotone{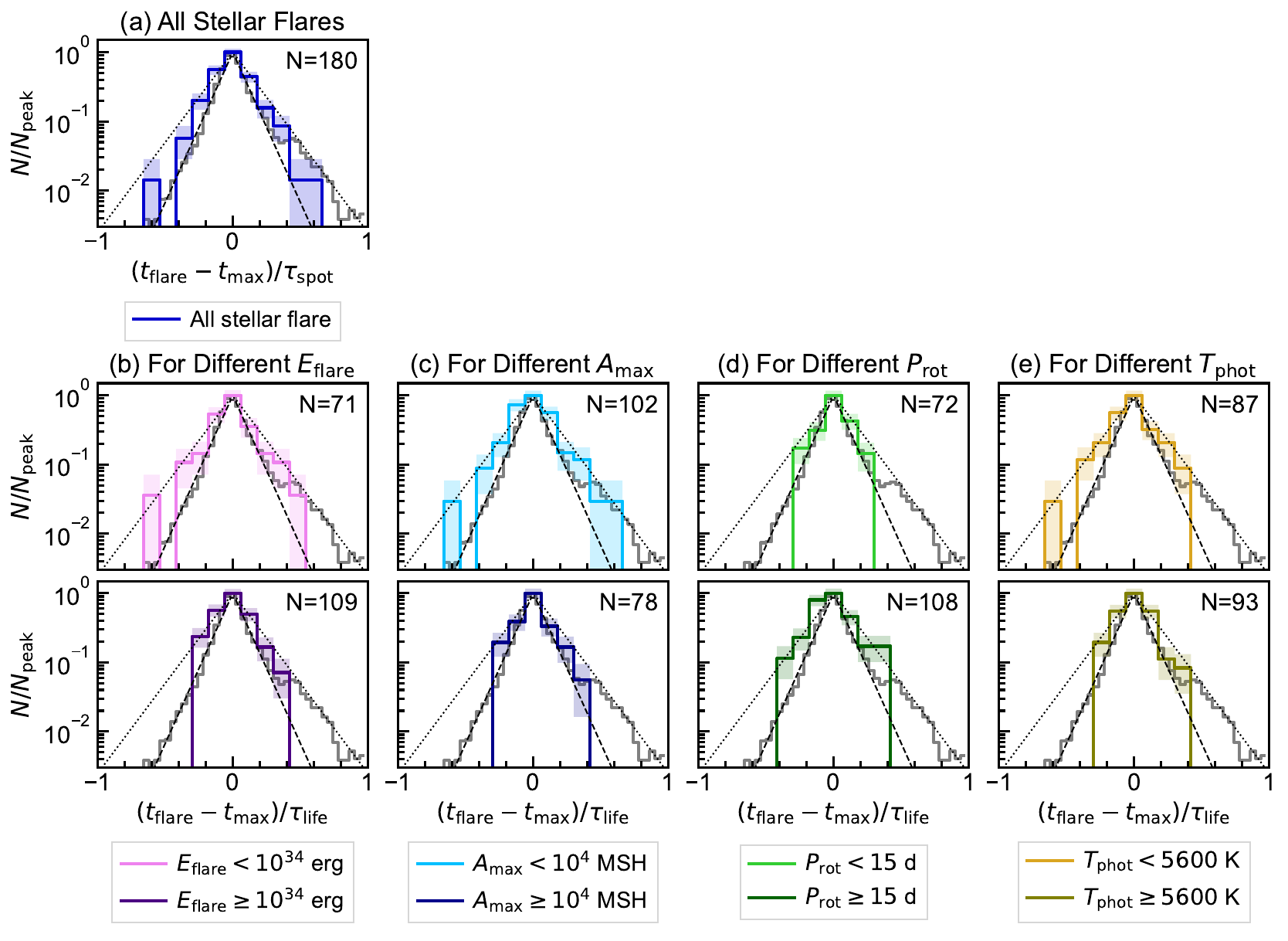}
\caption{
The same figure as Figure \ref{fig:flare_timing_solar} but for the stellar flares. The abscissas, ordinates, gray lines, black dots, and black dashed lines are all the same as in Figure \ref{fig:flare_timing_solar}. The colored line and color-shaded region in each panel show the frequency distribution normalized at $t_\mathrm{flare}-t_\mathrm{max}=0$ and its 1-$\sigma$ error for the stellar flares with sampling described below: panel (a) is for whole stellar flares; panel (b) is for the stellar flares sampled by different $E_\mathrm{flare}$ (top: $E_\mathrm{flare} < 10^{34} \, \mathrm{erg}$; bottom: $E_\mathrm{flare} \geq 10^{34} \, \mathrm{erg}$); panel (c) is for the stellar flares sampled by different $A_\mathrm{max}$ (top: $A_\mathrm{spot} < 10^{4} \, \mathrm{MSH}$; bottom: $A_\mathrm{spot} \geq 10^{4} \, \mathrm{MSH}$); panel (d) is for the stellar flares with different $P_\mathrm{rot}$ of the source stars (top: $P_\mathrm{rot} < 15 \, \mathrm{d}$; bottom: $P_\mathrm{rot} \geq 15 \, \mathrm{d}$); and panel (e) is for the stellar flares with different $T_\mathrm{phot}$ of the source stars (top: $T_\mathrm{phot} < 5600 \, \mathrm{K}$; bottom: $T_\mathrm{phot} \geq 5600 \, \mathrm{K}$).
\label{fig:flare_timing_stellar}}
\end{figure*}

In this section, we examine the property of the occurrence frequency distribution with respect to $(t_\mathrm{flare}-t_\mathrm{max})/\tau_\mathrm{life}$ both for solar flares and for stellar flares. These quantity can be paraphrased as the flare occurrence rate generated from each stage of sunspot evolution. Here, we show the histograms by normalizing the occurrence frequency of flares per year for each bin ($N$) with the that of the bin containing $t_\mathrm{flare}-t_\mathrm{max}=0$ ($N_\mathrm{peak}$), which is considered to take the maximum and to be less affected by the recurrent spots (see Section \ref{subsec:result_scatter}). 

Figure \ref{fig:flare_timing_solar} shows the frequency distribution of solar flare occurrence times for samples separated by its magnitudes and $A_\mathrm{max}$ of its source spot. When the source spots are similar in size, it is considered that the timing of when they turn into the backside or into the front side again is similar for the larger spot ($A_\mathrm{max} \geq 300$ MSH). Therefore, we can estimate the trend of recurrent spots by considering the maxima other than near the center. We note, however, that over- or underestimation due to false positives or false negatives of recurrent spot detection is expected (see also Section \ref{subsec:discussion_uncertainty}).

In light of this, we found that the flare occurrence rate through the time evolution of sunspots shows an analog shape regardless of the spot size and the flare magnitudes. More specifically, most histograms are roughly fitted by the exponential function:
\begin{align}
    \frac{N(t_\mathrm{flare})}{N_\mathrm{peak}} =    \begin{cases}
      & \mathrm{exp}\left(-\alpha_\mathrm{e} \dfrac{ |t_\mathrm{flare}-t_\mathrm{max}|}{\tau_\mathrm{life}} \right)  \\ 
      & \hspace{10pt} \textrm{ if } t_\mathrm{flare}-t_\mathrm{max} \leq 0 \\
      & \mathrm{exp}\left(-\alpha_\mathrm{d} \dfrac{ |t_\mathrm{flare}-t_\mathrm{max}|}{\tau_\mathrm{life}} \right) \\ 
      & \hspace{10pt} \textrm{ if } t_\mathrm{flare}-t_\mathrm{max} \geq 0
   \end{cases},
    \label{eq:reference_freq}
\end{align}
where $\alpha_\mathrm{e}$ and $\alpha_\mathrm{d}$ represent the index governing the steepness in the emergence and decaying phases, respectively. In this study, $\alpha_\mathrm{e,d}=6\textrm{-}10$ is suggested to fit (see dotted and dashed lines in Figure \ref{fig:flare_timing_solar}), while its value strongly depends on the model of $\tau_\mathrm{life}$ (see  Equations \ref{eq:tau_scaling} and \ref{eq:reference_freq}). The shape of this function shows the trend in which the flare occurrence rate takes its maximum when the sunspot takes its maximum area, and the flare occurrence rate decreases as the area of the sunspot is smaller. It means that the temporal variation of the flare occurrence rate is correlated with the time evolution of the spot area. (see Section \ref{subsec:result_correlation} in more detail).

It should be noted that the shape of histogram shown in Figure \ref{fig:flare_timing_solar} suggests a slightly asymmetric between the emergence and decaying phases, preferring $\alpha_\mathrm{e} \simeq 10$ and $\alpha_\mathrm{d} \simeq 6$. The relationship $\alpha_\mathrm{e} \geq \alpha_\mathrm{d}$ indicates that the distribution of flare occurrence rate is less likely to decrease with evolution during the decaying phase (see Equation \ref{eq:reference_freq}). Although it depends on the behavior near the peak, it suggests that the same or greater number of flares occur during the decaying phase as during the emergence phase. This result differs from the trend shown in previous studies that flares occur more frequently during the emergence phase \citep[][see Section \ref{subsec:discussion_interpret} in more detail]{Lee2012SoPh, Li2024ApJ}.  

Meanwhile, we find there are scattered behaviors that cannot be simply described by Equation (\ref{eq:reference_freq}). One is the trend that the larger the energy of the flare, the thinner the histogram (see the column-by-column differences in the same row in Figure \ref{fig:scatter_solar}). This is due to the fact that large flares are less likely to occur when sunspots are small, such as the beginning and last phases of spot evolution. Furthermore, the fact that relatively large sunspots are extracted when focusing on large flares, making the effect of sunspots turning to the backside more apparent. It may also contribute to this trend. The other is the flare with small $A_\mathrm{max}$ (see Figure \ref{fig:scatter_solar} with $A_\mathrm{max} \leq 100$ MSH), showing the moderate profiles in Figure \ref{fig:flare_timing_solar}. This is probably because the small spots have small $\mathrm{d}A(t)/\mathrm{d}t$, which causes inaccurate $t_\mathrm{max}$ owing to the accuracy of the sunspot data used in this study ($\sim 10$ MSH). Moreover, its error is reflected even via normalization due to a short $\tau_\mathrm{life}$ of small spots. 

Similarly to Figure \ref{fig:flare_timing_solar}, we show in Figure \ref{fig:flare_timing_stellar} the normalized frequency distribution of stellar flare occurrence times for the whole sample and for the sample with different flare, spot, and stellar properties. Specifically, panels (a), (b), (c), (d), and (e) respectively show the result for the whole stellar flares, the stellar flare with different $E_\mathrm{flare}$, the stellar flare with different $A_\mathrm{spot}$, the stellar flare with different $P_\mathrm{rot}$, and the stellar flare with different $T_\mathrm{phot}$. 

From panel (a), we found that the histogram for the whole sample is consistent with the model suggested by the solar case (Equation \ref{eq:reference_freq} with $\alpha_\mathrm{e,d}=6\textrm{-}10$). Moreover, the similarity of the top and bottom histograms in panels (a)-(d) indicates that its consistency is valid for the different parameters of flares (panel (b)), spots  (panel (c)), and stars (panels (d) and (e)). These results indicate that the correlation between flare occurrence rate and spot area shown in the solar flare is valid for stellar flare (see also Section \ref{subsec:result_correlation}). 

The asymmetry of flare occurrence rate between the emergence and decaying phases seen in the solar case is not clearly detected in the stellar flare, probably because the accuracy and number of samples for stellar flares are not sufficient (see Section \ref{subsec:discussion_uncertainty}).

\subsection{Correlation between Spot Area and Flare Occurrence Rate}
\label{subsec:result_correlation}

\begin{figure}[t!]
\centering
\plotone{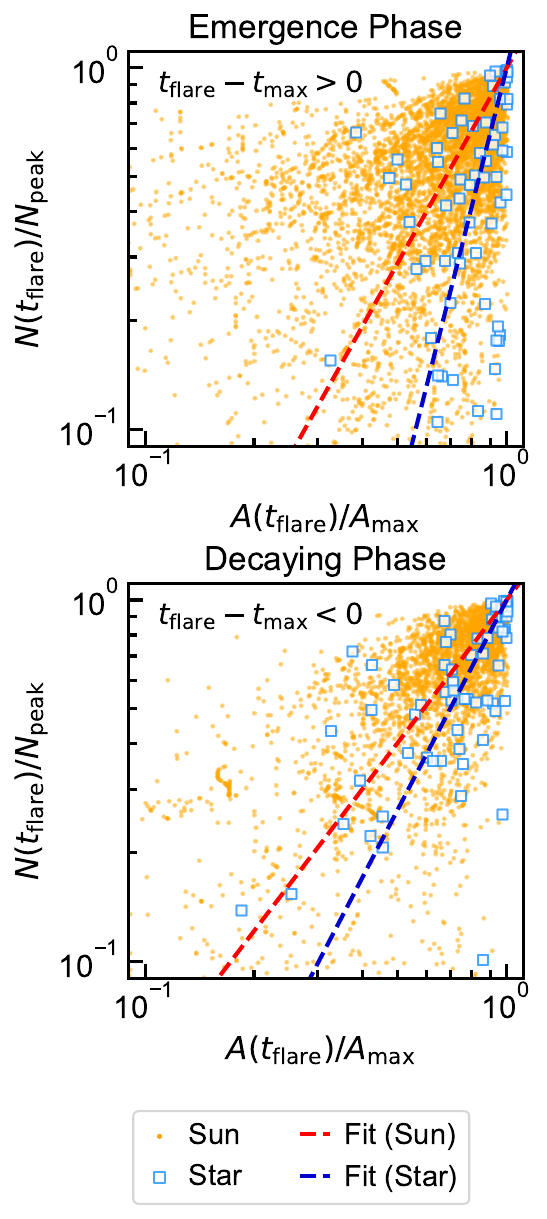}
\caption{Scatter plot of the normalized spot area at the flare occurrence time, $A(t_\mathrm{flare})/A_\mathrm{max}$, and the normalized flare occurrence rate, $N(t_\mathrm{flare})/N_\mathrm{peak}$ for each flare at the emergence phase (upper panel) and the decaying phase (lower panel). The orange points and blue squares correspond to solar and stellar flares, respectively. We calculate $N(t_\mathrm{flare})/N_\mathrm{peak}$ from $(t_\mathrm{flare}-t_\mathrm{max})/\tau_\mathrm{life}$ using Equation (\ref{eq:reference_freq}) with $\alpha_\mathrm{e}=10$ and $\alpha_\mathrm{d}=6$. The dashed lines are the fitted lines using Equation (\ref{eq:scaling_gamma}), showing the blue ones are the results of solar flares and the blue ones are the results of stellar flare. The median value in Table \ref{tab:index_spot_flarerate} was used as the index $\gamma$.
\label{fig:spot_flarerate}}
\end{figure}

\begin{table}[t!]
    \centering
    \caption{Power-law index $\gamma$ in Equation (\ref{eq:scaling_gamma}) derived by Figure \ref{fig:spot_flarerate}. }
    \begin{tabular}{ccc} 
	\hline
      & Emergence Phase & Decaying Phase \\
    \hline
    Solar Flare  & $\gamma=1.8^{+1.0}_{-0.7}$ & $\gamma=1.3^{+0.6}_{-0.4}$ \\
    Stellar Flare & $\gamma=4.0^{+2.4}_{-1.6}$ & $\gamma=1.9^{+1.3}_{-0.3}$  \\
    \hline
    \end{tabular} 
    \tablecomments{The power-law indexes are obtained by the median and 1-$\sigma$ range of the frequency distribution of $\log (N(t)/N_\mathrm{peak})/\log (A(t)/A_\mathrm{max})$.}
    \label{tab:index_spot_flarerate}
\end{table}

Section \ref{subsec:result_frequency} shows the correlation between the temporal variation of the flare rate ($N(t)/N_\mathrm{peak}$; see Equation \ref{eq:reference_freq}) and the time evolution of spot area $A(t)/A_\mathrm{max}$. A possible approach would be to fit the spot evolution as a function of $(t_\mathrm{flare}-t_\mathrm{max})/\tau_\mathrm{life}$, but in practice the large dispersion makes this difficult. Based on the above, we investigated this correlation using the scatter plots.

Figure \ref{fig:spot_flarerate} shows the scatter plots of $A(t)/A_\mathrm{max}$ and $N(t)/N_\mathrm{peak}$  for each solar flare (orange points) and stellar flare (blue squares) at the emergence phase (upper panel) and the decaying phase (lower panel). In the calculation of $N(t)/N_\mathrm{peak}$ from $(t_\mathrm{flare}-t_\mathrm{max})/\tau_\mathrm{life}$, we used Equation (\ref{eq:reference_freq}) with $\alpha_\mathrm{e}=10$ and $\alpha_\mathrm{d}=6$. Since a sunspot area near the solar limb is underestimated in the NOAA data, we only plot $A(t)/A_\mathrm{max}$ for the flares whose heliocentric longitudes are within $\pm 45^\circ$. Also, we calculated the power-law index $\gamma$ of the scaling relation 
\begin{align}
    \frac{N(t)}{N_\mathrm{peak}} \propto \left( \frac{A(t)}{A_\mathrm{max}} \right)^{\gamma}
    \label{eq:scaling_gamma}
\end{align}
by fitting these scatter plots, respectively (results are shown in Table \ref{tab:index_spot_flarerate}). The lines corresponding to the best-fitting value are written as dashed lines in Figure \ref{fig:spot_flarerate}. 

In this figure, we found the trend that the larger the spot is in its evolutionary stage, the more flares are likely to occur, in both solar and stellar cases. However, due to a large dispersion, it is difficult to quantify the correlation. Even for the Sun, which has a large sample, there is a large uncertainty of $\gamma$ values, which is about 50 percent of relative error. It is larger in the stellar case due to the small sample size. As a result, we can only suggest $\gamma = 1\textrm{-}3$, which is a less severe restriction. In this respect, \citet{Maehara2017PASJ} shows that the flare occurrence rate is roughly proportional to the spot area at the time of flare occurrence, $N(t_\mathrm{flare}) \propto A(t_\mathrm{flare})$, leading to $\gamma = 1$.

\subsection{Nature of Flares at Different Stages of Spot Evolution}
\label{subsec:result_property}

\begin{figure*}[t!]
\centering
\plotone{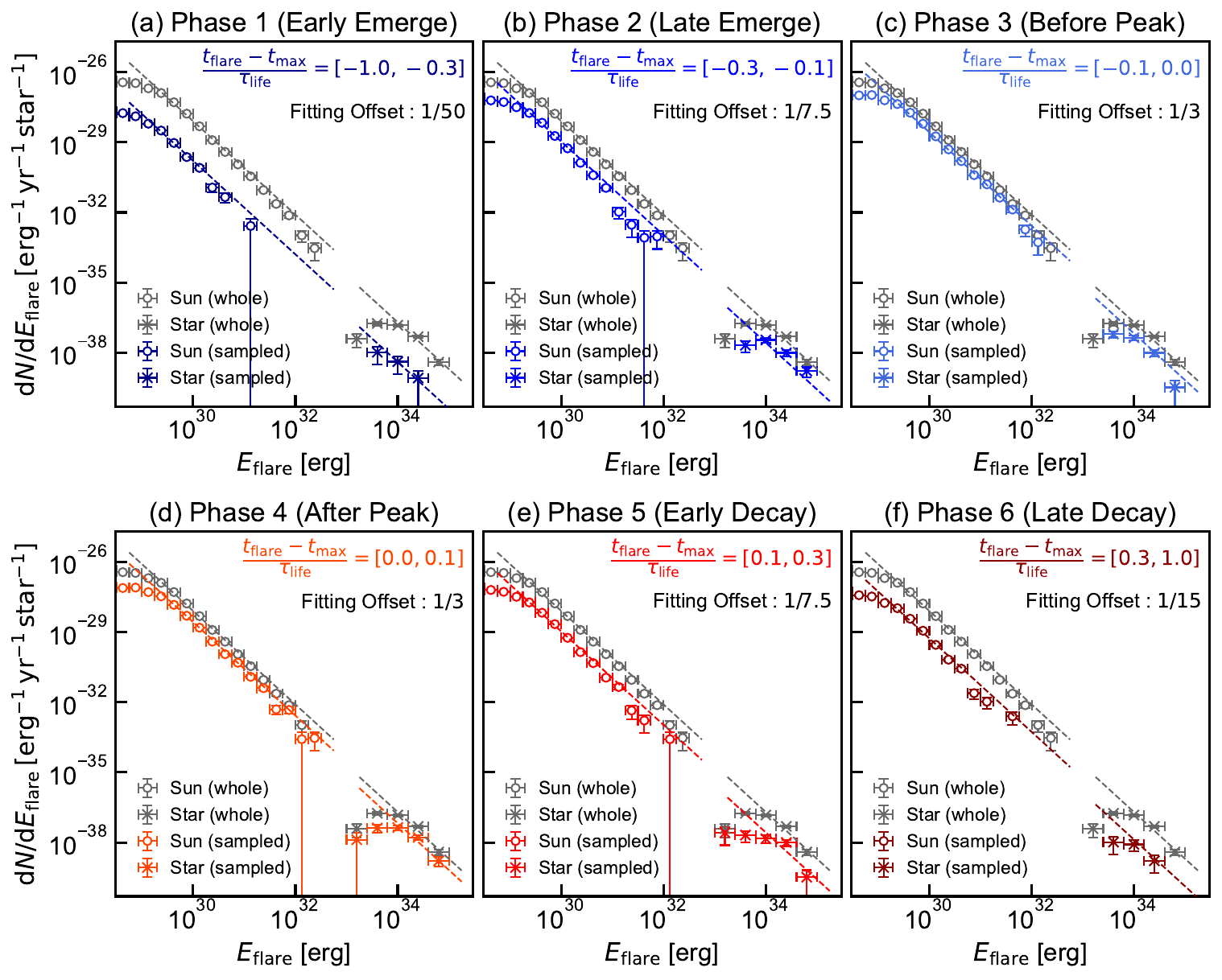}
\caption{Occurrence frequency distribution per energy of solar flares and stellar flares on solar-type stars with different spot evolutionary phases. 
Panels (a)-(f) show the result of the sample for each stage of spot evolution, respectively showing $(t_\mathrm{flare}-t_\mathrm{max})/\tau_\mathrm{life}=[-1.0, -0.3]$, $[-0.3, -0.1]$, $[-0.3, -0.1]$,  $[-0.1, 0.0]$,  $[0.0, 0.1]$,  $[0.1, 0.3]$, and  $[0.3, 1.0]$. The horizontal and vertical axis of each panel indicates the flare energy ($E_\mathrm{flare})$ and the flare occurrences rate per star per year and per unit energy ($\mathrm{d} N/\mathrm{d} E_\mathrm{flare})$, respectively. $\mathrm{d} N/\mathrm{d} E_\mathrm{flare}$ is calculated by dividing the number of flares in each phase and each energy bin by the observation time for all phases. We show the result of solar flares by the circle symbols and the result of stellar flares by the cross symbols, with the error bars representing 1-$\sigma$ uncertainty. The color of the symbol corresponds to the sample category: the gray one is the whole sample, and the colored one is the extracted sample. Dotted lines in each panel show $\mathrm{d}N/\mathrm{d}E \propto E^{-2}$, with the gray line representing the fitting of the whole sample and the colored line representing the fitting of the extracted sample. The offset between the two lines is shown in the upper right corner.
\label{fig:energy_distribution}}
\end{figure*}

\begin{figure*}[t!]
\centering
\plotone{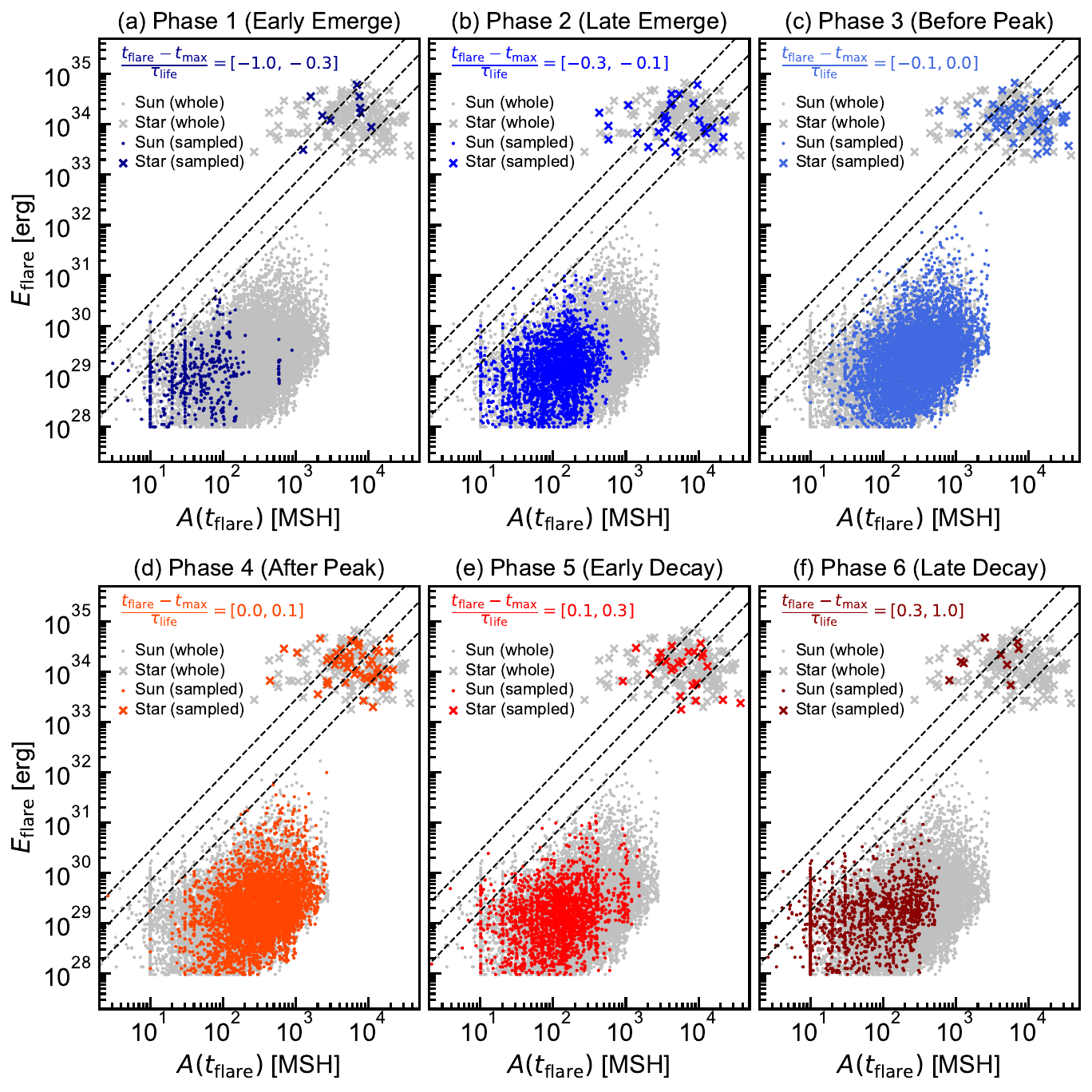}
\caption{Scatter plot of the flare energy ($E_\mathrm{flare}$) and spot area at the flare occurrence ($A(t_\mathrm{flare})$) of solar flares and stellar flare with different spot evolutionary phases. The criterion for sample extraction and panel arrangement are the same as in Figure \ref{fig:energy_distribution}. The dotted lines represent the upper limit of $E_\mathrm{flare}$ determined by the spot area with $f=0.1$ and $B= 500, 1000, 2000 \, \mathrm{G}$ from the bottom one to top one, where $f$ is the fraction of energy released by a flare, and $B$ are the magnetic field strength of the spot \citep[see][]{Shibata2013PASJ, Maehara2015EPS}.
\label{fig:spot_energy}}
\end{figure*}

The results of Section \ref{subsec:result_frequency} suggest that the flare occurrence rate have a similar $E_\mathrm{flare}$ dependence in the overall sample and in the sample extracted in a certain range of $(t_\mathrm{flare}-t_\mathrm{max})/\tau_\mathrm{life}$. To confirm this, we investigated the validity of $E_\mathrm{flare}$ dependence of the flare occurrence rate,  $\mathrm{d}N/\mathrm{d}E_\mathrm{flare} \propto E_\mathrm{flare}^{-2}$ \citep{Crosby1993SoPh, Shimizu1995PASJ, Maehara2012Natur, Shibayama2013ApJS, Notsu2019ApJ}, when flares that occurred at different sunspot evolution stages were extracted. Here, we defined $(t_\mathrm{flare}-t_\mathrm{max})/\tau_\mathrm{life}=$ $[-1.0, -0.3]$, $[-0.3, -0.1]$, $[-0.3, -0.1]$,  $[-0.1, 0.0]$,  $[0.0, 0.1]$,  $[0.1, 0.3]$, and  $[0.3, 1.0]$ as phases 1-6 \footnote{Phase 1 and 2 correspond to the early and late part of the growth phase, Phase 3 to just before the peak, Phase 4 to just after the peak, and Phase 5 and 6 to the early and late part of the decline phase, respectively.}, respectively. By dividing the number of flares in each phase and each energy bin by the observation time for all phases, i.e., 28.75 years for the Sun and 4 years for solar-type stars, we calculated $\mathrm{d}N/\mathrm{d}E_\mathrm{flare}$ for both the solar and stellar flares. For calculation, we conducted the correction based on gyrochronology for the normalization by stellar number with $P_\mathrm{rot} \geq 10 \, \mathrm{d}$ \citep[20672 stars; see][]{Maehara2017PASJ,Okamoto2021ApJ}. We also conduct the corrections with respect to the detection probability of source spots both for solar and stellar samples (see Table \ref{tab:number_of_flares}). 

The result is shown in Figure \ref{fig:energy_distribution}. The values for each phase (colored symbols in each panel) are lower than the values using the whole sample because we use the total observation time for the calculation of $\mathrm{d}N/\mathrm{d}E_\mathrm{flare}$ in all cases. Also, we can confirm that the most dominant values are near the peak, as discussed in Section \ref{subsec:result_frequency}. From this figure, we can confirm that the scaling of $\mathrm{d}N/\mathrm{d}E_\mathrm{flare} \propto E_\mathrm{flare}^{-2}$ fits the extracted solar sample as well as the whole solar sample. The same can be confirmed for stellar flare. Additionally, the results for the extracted stellar sample can be fitted using the same offset from the whole sample with some slight deviations. 

Using the same sample classification, we can examine different characteristics. Figure \ref{fig:spot_energy} shows a scatter plot of the spot area at the flare occurrence $A(t_\mathrm{flare})$ and flare energy $E_\mathrm{flare}$\footnote{We note the similar scatter plots in \citet{Shibayama2013ApJS} and \citet{Okamoto2021ApJ} are different from Figure \ref{fig:spot_energy} in that they take $A_\mathrm{max}$ as the horizontal axis, while Figure \ref{fig:spot_energy} takes $A(t_\mathrm{flare})$.}. As in Section \ref{subsec:result_correlation}, only the flares whose heliocentric longitudes are within $\pm 45^\circ$ are considered. It displays that the scattering of the solar flare moved in the direction of $A(t_\mathrm{flare})$ as the phase progressed, fulfilling the upper limit of $E_\mathrm{flare}$ determined by the spot area \citep{Shibata2013PASJ, Maehara2015EPS}. This behavior can be explained by the property that larger spots are more likely to cause strong flares and that the evolution of the spot is analogous (see Section \ref{subsec:result_scatter} and Section \ref{subsec:discussion_interpret}). 

We note that the trend is not clearly visible in the stellar case, probably due to the small sample size, the uncertainty of $A(t_\mathrm{flare})$ (see Section \ref{subsec:discussion_uncertainty}), and the diversity of spot property.

\section{Discussion}
\label{sec:discussion}

\subsection{Temporal Variation of Flare Occurrence Rate per Spot}
\label{subsec:discussion_time_variation}

This section provides a description of the temporal variation of flare occurrence rate due to spot evolution based on the result in Section \ref{sec:result}. Specifically, we formulate the flare occurrence rate per spot given the flare energy $E_\mathrm{flare}$ and the maximum area $A_\mathrm{max}$, describing it as $\mathcal{N}(t,A_\mathrm{max},E_\mathrm{flare})$ hereafter.

It should be emphasized that the flare occurrence rate in Equation (\ref{eq:reference_freq}), $N$, is not per spot but per star. Nevertheless, we can assume that this equation is valid for each spot because of the independence of $A_\mathrm{max}$ and $E_\mathrm{flare}$. Thus, we can obtain from Equation (\ref{eq:reference_freq})
\begin{align}
    & \mathcal{N}(t,A_\mathrm{max},E_\mathrm{flare}) = \mathcal{N}(t=t_\mathrm{max}, A_\mathrm{max},E_\mathrm{flare}) \times \notag \\
    & 
     \begin{cases}
      \mathrm{exp}\left(-\alpha_\mathrm{e} \dfrac{ |t-t_\mathrm{max}|}{\tau_\mathrm{life}} \right) & \textrm{ if } t-t_\mathrm{max} \leq 0 \\
      \mathrm{exp}\left(-\alpha_\mathrm{d} \dfrac{ |t-t_\mathrm{max}|}{\tau_\mathrm{life}} \right) & \textrm{ if } t-t_\mathrm{max} \geq 0
   \end{cases},
    \label{eq:N_t_dependence}
\end{align}
where $\mathcal{N}(t=t_\mathrm{max}, A_\mathrm{max},E_\mathrm{flare})$ is the flare occurrence rate when the sunspot takes the maximum area $A_\mathrm{max}$. If we use Equation (\ref{eq:scaling_gamma})  instead of Equation (\ref{eq:reference_freq}), we can obtain another expression as
\begin{align}
    \frac{\mathcal{N}(t,A_\mathrm{max},E_\mathrm{flare})}{\mathcal{N}(t=t_\mathrm{max}, A_\mathrm{max},E_\mathrm{flare})} =  \left( \frac{A(t)}{A_\mathrm{max}} \right)^\gamma,
    \label{eq:N_t_dependence_v2}
\end{align}
while $\gamma$ includes the uncertainty as shown in Section \ref{subsec:result_correlation}.

In addition, \citet{Maehara2017PASJ} analyze the flare occurrence rate for different spot areas and shows that the relation  $\mathrm{d}\mathcal{N}/\mathrm{d}E_\mathrm{flare} \propto E_\mathrm{flare}^{-2}$ holds regardless of $A(t_\mathrm{flare})$ and its coefficient is roughly proportional to $A(t_\mathrm{flare})$. On this basis, we roughly derive the relation as
\begin{gather}
    \mathcal{N}(t=t_\mathrm{max}, A_\mathrm{max},E_\mathrm{flare}) \propto A_\mathrm{max}E_\mathrm{flare}^{-1}. \label{eq:N_AandE_dependence}
\end{gather}

Consequently, the formulation consisting of Equations (\ref{eq:N_t_dependence}) and (\ref{eq:N_AandE_dependence}) (or Equations (\ref{eq:N_t_dependence_v2}) and (\ref{eq:N_AandE_dependence})) provides dependence of $\mathcal{N}(t,A_\mathrm{max},E_\mathrm{flare})$ on $t$, $A_\mathrm{max}$ and $E_\mathrm{flare}$. This formulation can be said to be an extension of the flare occurrence rate per star, which is investigated by previous studies. We consider it provides a benchmark for predicting when and how much flares will occur from a sunspot/starspot. To obtain a more detailed profile, it is possible to incorporate detailed properties such as the fact that large flares are less likely to occur from small spots \citep[][see also Figure \ref{fig:spot_energy}]{Shibata2013PASJ, Maehara2015EPS}.

As a supplement, we check for consistency with the result of this paper. $\mathrm{d}N/\mathrm{d}E_\mathrm{flare}$ in Section \ref{subsec:result_property} can be derived by the time-integration to choose spot evolutionary stages and the convolution under the appearance frequency of spot per year with $A_\mathrm{max}$ against $\mathrm{d} \mathcal{N}(t, A_\mathrm{max},E_\mathrm{flare}) /{\mathrm{d}E_\mathrm{flare}}$. Thus, using Equations (\ref{eq:N_t_dependence}) and (\ref{eq:N_AandE_dependence}), it is easy to see that $\mathrm{d}N/\mathrm{d}E_\mathrm{flare}$ maintains the dependence of $\propto E_\mathrm{flare}^{-2}$ for any time-integral range. This result is consistent with the result in Figure \ref{fig:energy_distribution}.

\subsection{Implications of Trends}
\label{subsec:discussion_interpret}

Sections \ref{subsec:result_frequency} and \ref{subsec:result_correlation} show that if we focus on a particular spot, the larger the spot area increases with time evolution, the more frequently flares occur. This suggests that the near-surface magnetic energy stored in the spot region plays an essential role to cause reconnection events that triggers flares. This interpretation is also valid in explaining the correlation between the flare energy and the spot area \citep[e.g.,][see also Figure \ref{fig:spot_energy}]{Sammis2000ApJ, Shibata2013PASJ, Okamoto2021ApJ} or between the flare frequency and the spot area \citep[e.g.][]{Giovanelli1939ApJ, Greatrix1963MNRAS, Maehara2017PASJ, Okamoto2021ApJ, Tu2021ApJS}. In this respect, our study shows that this interpretation shown in previous studies can be extended in the temporal direction, i.e., to all stages of spot evolution.

The fact that similar behavior was observed for the Sun and solar-type stars suggests the common nature of flare-productive starspot candidates in them. Previous studies have shown that spots and flares, respectively, have several common trends in the Sun and solar-type stars: the energy distribution of flares \citep{Maehara2012Natur, Shibayama2013ApJS, Maehara2015EPS, Tu2020ApJ, Okamoto2021ApJ, Vasilyev2024Sci}; the correlation between of flare energy and duration \citep{Namekata2017ApJ, Maehara2017PASJ}; and the timescale of spot evolution for spots \citep{Namekata2019ApJ, Namekata2020ApJ}. Combined with these results, the commonality shown in this work enhances the possibility that the spot-to-flare magnetic activity on solar-type stars is caused by similar physical processes to those on the Sun.

In addition to that, the results of Section \ref{subsec:result_frequency} suggest that the occurrence rate of flares during the decaying phase matches or exceeds that during the emergence phase both for the solar and stellar cases. This tendency differs from previous suggestions that have shown that flares are more likely to occur during the emergence phase in the Sun \citep{Lee2012SoPh,Li2024ApJ}. We attribute this difference to our treatment of recurrent spots (Sections \ref{subsec:method_solar} and \ref{subsec:method_stellar}), normalization (Section \ref{subsec:method_normalize}), and quantitative treatment of flare timing (Section \ref{sec:result}), which have not been considered in previous studies. Thus, this study provides an opportunity to focus renewed attention on the flare occurrence during the decaying phase, which has been underestimated.

From a physical point of view, the magnetic reconnection, which triggers flares, can be interpreted as the release of the magnetic free energy at sunspots with complex magnetic field structures \citep[e.g.,][]{Kunzel1960AN,Bornmann1994SoPh, Toriumi2017ApJ, ToriumiTakasao2017ApJ}. In this regard, the flux emergence associated with sunspot formation has been considered the key process to create the magnetic free energy \citep[e.g.,][]{Archontis2008JGRA, Cheung2014LRSP}. However, based on the results of this study, the magnetic free energy due to the cancellation of magnetic flux during the decaying phase may be important \citep[e.g.,][]{Martin1985AuJPh,vanBallegooijen1989ApJ, Welsch2006ApJ}. We consider that the event analyses and magnetohydrodynamic simulations of flares occurring during the decaying phase are expected to help interpret this trend.

\subsection{Uncertainties and Robustness}
\label{subsec:discussion_uncertainty}

\begin{figure}[t!]
\centering
\plotone{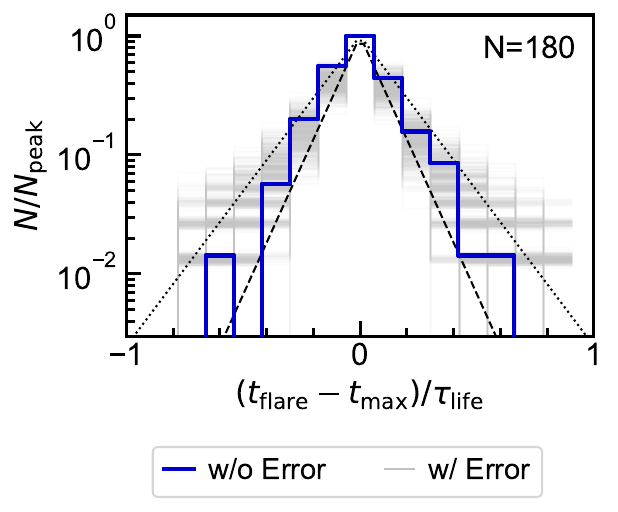}
\caption{
Comparison between the histogram using the observed stellar data (blue line, same as Figure \ref{fig:flare_timing_stellar} (a)) and the histograms using the stellar data to which the uncertainties induced by the local minima tracing method are artificially added (gray lines). For the latter, the results for 1000 patterns are shown.  The abscissas, ordinates, black dots, and black dashed lines are all the same as in Figure \ref{fig:flare_timing_solar}.
\label{fig:flare_timing_stellar_robust}}
\end{figure}

Although our study shows certain results, 
it is true that our procedure for obtaining $(t_\mathrm{flare}-t_\mathrm{max})/\tau_\mathrm{life}$ involves some uncertainties. The uncertainties of $\tau_\mathrm{life}$ and $t_\mathrm{flare}-t_\mathrm{max}$ each could impact the results, particularly when the number of data points is small (e.g., Figure \ref{fig:flare_timing_stellar}). Here, we discuss each of these uncertainties and the robustness of the results in this work.

The most significant simplification that could affect the results is the normalization using Equation (\ref{eq:tau_scaling}). In the stellar case, estimating $A_\mathrm{max}$ via the local-minima tracing method has limitations \citep[see also][]{Namekata2019ApJ, Namekata2020ApJ}. This method assumes the darkening from the largest spots is dominant, but if multiple similar-sized spots exist nearby, $A_\mathrm{max}$ is overestimated due to their indiscernibility. Conversely, spots away from the disk center appear smaller due to line-of-sight projection effects and show weaker photometric variation due to the limb-darkening effect, both of which cause underestimation. It is also possible that the accurate estimation of $\Delta F/F$ is prevented by faculae (see Equation \ref{eq:A-Tspot}). These effects introduce uncertainty in spot lifetime estimates using Equation (\ref{eq:tau_scaling}). Also, this uncertainty propagates to the estimated starspot lifetimes for solar-type stars in Figure \ref{fig:lifetime} (see Appendix \ref{app:scaling_tau}). Additionally given that the solar data in Figure \ref{fig:lifetime} show large dispersion, the validity of Equation (\ref{eq:tau_scaling}) is uncertain. In order to obtain a more precise formulation, it is important to improve the estimation accuracy of $A_\mathrm{max}$, for example, by taking inclination effects into account \citep[e.g.,][]{Notsu2019ApJ} or by applying detailed spot mapping techniques (see the next paragraph). In addition, increasing the sample size of starspots is also necessary to enhance statistical robustness (see also Section \ref{subsec:summary}).

Similarly, $t_\mathrm{flare}-t_\mathrm{max}$ has uncertainties. For the Sun, $t_\mathrm{max}$ may be inaccurate due to the existence of sunspots that have taken $A_\mathrm{max}$ on the backside or the false identification of recurrent sunspots due to our simplified method (see Appendix \ref{app:recurrent_sunspot}). In the stellar case, A is expected to have an uncertainty due to the lack of sensitivity of the local minima trace method with respect to phase (see Appendix \ref{app:method_sun-as-a-star}). Additionally, the uncertainties of $A_\mathrm{max}$ discussed in the previous paragraphs may lead to misidentification for cycles that take the maximum value.
Moreover, misidentification of flare-productive spots cannot be rejected when a flare originates from an unresolved small spot or a polar spot that is constantly visible. The frequency of such misidentification warrants future investigation using the (Zeeman) Doppler imaging technique \citep{Vogt1983PASP, Semel1989A&A, Donati2006MNRAS} or advanced light curve modeling \citep{Strassmeier1992A&A, Croll2006PASP, Ikuta2020ApJ, Ikuta2023ApJ}.

In light of the above discussion, it is crucial to evaluate the robustness of our results in the stellar case, which is considered to contain more uncertainty. We tested this by introducing mock errors in the stellar data. Specifically, we examined the impact of uncertainties in $A_\mathrm{max}$ and $t_\mathrm{max}$ caused by local-minima tracing methods. The range of uncertainty is based on sun-as-a-star analysis (see Appendix \ref{app:method_sun-as-a-star}). By scaling $A_\mathrm{max}$ with a log-uniform random factor $[1/4, 4]$ and shifting $t_\mathrm{max}$ by a uniform random value in $[-P_\mathrm{rot}/4, P_\mathrm{rot}/4]$, we generated 1000 mock datasets. Figure \ref{fig:flare_timing_stellar_robust} compares their histograms (gray lines) with the histogram shown in Figure \ref{fig:flare_timing_stellar} (blue line). Uncertainty distorts the histograms, but less than 1 percent of cases are inconsistent from Equation (\ref{eq:reference_freq}) with $\alpha_\mathrm{e,d}=6\textrm{-}10$, including uncertainty in sample size. This is likely because the dominant samples satisfying $t_\mathrm{flare} - t_\mathrm{max} \sim 0$ are minimally affected. Thus, we conclude that the result in Section \ref{subsec:result_frequency} remains statistically robust within these uncertainties. However, we also find that these uncertainties hinder precise determination of $\alpha_\mathrm{e,d}$ with the current sample size.

\section{Summary and Future prospect}
\label{subsec:summary}

The spot-to-flare activity of the Sun and solar-type stars has attracted much attention in terms of solar/stellar physics and planetary science. Although it is expected that there is a relationship between the nature of sunspots and flares, the analysis considering time-directional information has rarely been investigated. The aim of our work is to investigate how the flare rate changes in response to spot evolution and to compare these trends between the Sun and solar-type stars. For this purpose, we have conducted two analyses: to compile the archive data of sunspots and solar flares observed by NOAA and GOES for the Sun and to conduct the time-series analysis of the \textit{Kepler} light curves, which follows the local-minima tracing method \citep{Namekata2019ApJ,Namekata2020ApJ}, to track the time evolution of flare-productive starspot candidates with the stellar flare reported by \citet{Okamoto2021ApJ} for solar-type stars. As a result, we have obtained the flare timing during spot evolution ($t_\mathrm{flare}-t_\mathrm{max}$) for each of the obtained 24124 solar flares and 180 stellar flares. By applying a normalization using spot lifetime ($\tau_\mathrm{life}$), we have statistically analyzed the scale-independent property of flare occurrence rates at each spot evolution stage.

The main results in this study are as follows: (i) The flare occurrence distribution for $(t_\mathrm{flare} - t_\mathrm{max})/\tau_\mathrm{life}$ shows a similar distribution regardless of spot size, flare energy, or stellar property. Using this, we derive the scale-independent description of temporal variation of flare occurrence rate per spot, which is common to the Sun and solar-type stars (Equations \ref{eq:N_t_dependence} and \ref{eq:N_AandE_dependence}); (ii) The flare occurrence rate shows temporal variation linked to spot evolution, increasing as the sunspot emerges during the emergence phase and decreasing as the sunspot decays during the decaying phase. This means that the amount of magnetic field energy due to spot regions is an important factor in the flare occurrence process; (iii) The flare occurrence rates during the decaying phase may not differ from the emergence phase or, in the case of the Sun, may be higher. This suggests the importance of reconnection events caused by the diffusion of magnetic fields. This is a trend that was first seen through quantitative analysis of flare timing in this study. These results are benchmarks of the nature of flare-productive spots in the Sun and solar-type stars, promoting more detailed studies to examine the trends suggested in this study.

Besides, the relationship between flare occurrence frequency and stellar/spot properties, which was not covered in this study, is also a subject that deserves future investigation. Solar observations suggest that complex magnetic structures influence flare frequency \citep{Sammis2000ApJ, Maehara2017PASJ}, raising the question of whether such a trend exists among starspots. For example, a difference in spot evolution may exist between flare-productive/non-flare-productive starspots. Although our simple framework did not detect a clear distinction, a more quantitative discussion would require a detailed analysis. Additionally, some stars with similar-sized spots exhibit superflares repeatedly, while others do not \citep{Shibayama2013ApJS, Okamoto2021ApJ}. In order to investigate the cause of this discrepancy, it is necessary to characterize each solar-type star in more detail.

Further observations of the magnetic activity of solar-type stars could help investigate the cause of this discrepancy. In this regard, the PLAnetary Transits and Oscillations of stars (PLATO) mission \citep{Rauer2014ExA, Rauer2024arXiv, Breton2024A&A} may provide the opportunity to address it, expanding the sample of flare-productive starspot candidates in solar-type stars and constraining rotational profiles of them via photometric/asteroseismic analysis \citep{Gizon2003ApJ,Gizon2004SoPh,Kamiaka2018MNRAS, Benomar2018Sci}. Also, It is important to increase the detailed event analysis of spot-to-flare activity via multi-wavelength or multi-technique observational campaigns  \citep[e.g.,][]{Namekata2024aApJ, Namekata2024bApJ}.

\section*{Acknowledgment} \label{sec:ackn}
This paper includes data collected by the \textit{Kepler} mission. Funding for the \textit{Kepler} mission is provided by the NASA Science Mission Directorate. The \textit{Kepler} data presented in this paper were obtained from the Multimission Archive at STScI.

We acknowledge with thanks for T.K. Suzuki and K. Kakiuchi for insightful discussion for this study. We also thank K. Petrovay for kindly providing us their data. This work was supported by JSPS
(Japan Society for the Promotion of Science) KAKENHI Grant Numbers JP24KJ0605 (T.T.), JP21J00316 (K.N.), JP24K00680 (K.N., H.M.), JP24H00248 (K.N., H.M.), JP20H05643 (M.H.), JP20KK0072 (S.T.), JP21H01124 (S.T.) and JP21H04492 (S.T.).


\appendix

\section{Detection of Recurrent Sunspot}
\label{app:recurrent_sunspot}

This section describes the details of our procedure to detect recurrent sunspots using sunspot position and area. First, we detected recurrent sunspot candidates using the sunspot's position. Among the pairs of a sunspot that turn to the backside and a sunspot that come out the front side, those with a difference in the latitude within 4 degrees and the Carrington longitude within 8 degrees were selected as a recurrent sunspot candidate. These thresholds were determined from Figure 4 of \citet{Henwood2010SoPh}. We note that the longitude was corrected for the expected shift from differential rotation at that latitude. Also, given that some sunspots rotate slowly \citep{Nagovitsyn2018AstL}, we extracted the candidates using both the differential rotation profile from \citet{Snodgrass1990ApJ} and a profile slowed by 0.2 degrees per day from their one.

Next, for the obtained candidates, we extract the recurrent sunspots by using the sunspot area. It is suggested that the spot evolution is formulated as $\mathrm{d} A(t)/\mathrm{d} t \propto A(t)^{\beta}$, where $\beta$ is the scaling index. Although there are various suggestions for the value of $\beta$ from $\beta=0$ \citep{Gnevyshev1938IzPul,Waldmeier1955,Henwood2010SoPh} to $\beta=0.5$ \citep{MartinezPillet1993A&A}, we used $\beta=0.3$ as suggested in \citet{Namekata2019ApJ, Namekata2020ApJ}. By solving this, we obtain the spot evolution as
\begin{align}
    \dfrac{A(t)}{A_\mathrm{max}} = 
    \begin{cases}
        \left( 1 + \dfrac{t-t_\mathrm{max}}{\tau_\mathrm{emerge} } \right)^{1/(1-\beta)} \ \textrm{if} \ t \leq t_\mathrm{max}  \\
        \left( 1 - \dfrac{t-t_\mathrm{max}}{\tau_\mathrm{decay} } \right)^{1/(1-\beta)} \ \textrm{if} \ t \geq t_\mathrm{max}
    \end{cases}, \label{eq:app_spot_evolution}
\end{align}
where $\tau_\mathrm{emerge}$ and $\tau_\mathrm{decay}$ are the duration satisfying $A(t_\mathrm{max}-\tau_\mathrm{emerge})=0$ in the emergence phase and $A(t_\mathrm{max}+\tau_\mathrm{decay})=0$ in the decaying phase, respectively. According to \citet{Namekata2019ApJ}, $|\mathrm{d} A(t)/\mathrm{d} t|$ differs roughly by a factor of 2 between the emergence and decaying phases, suggesting that $\tau_\mathrm{decay} = 2\tau_\mathrm{emerge}$ holds. Thus, because of $\tau_\mathrm{life}=\tau_\mathrm{emerge}+\tau_\mathrm{decay}$, Equation (\ref{eq:app_spot_evolution}) is determined when $\tau_\mathrm{life}$ is given.
Here, we considered tau according to Equation (\ref{eq:tau_scaling}) to be the standard spot lifetime, $\tau_\mathrm{life, std}$. In order to also take into account the dispersion of age per sunspot, a candidate whose spot area evolves between the model with $\tau_\mathrm{life, std}/5$ and the model with $5\tau_\mathrm{life, std}$ was considered as a recurrent sunspot.

\section{Setting of Clustering}
\label{app:Clustering}

In this section, we introduce the detailed procedure for clustering local minima where the difference in the rotational phase is within 0.10 in adjacent rotational cycles. As indicated in Section \ref{subsec:method_stellar}, we conduct the \texttt{DBSCAN} clustering for the local minima in the two-dimensional space $(m, C\ell)$ space with $n_\mathrm{cluster}=2$, where $m$ and $\ell$ is the rotational cycle and phase, respectively. The analysis is performed under the periodic boundary condition $(m,C\ell)=(m\pm1,C(\ell \mp1 ))$, which holds by the definition of $m$ and $\ell$ (see Section \ref{subsec:method_stellar}).

Here, we set the parameters governing the clustering as $ C = \sqrt{3}/\delta $ and $ \varepsilon = 2$, where $\delta$ is the threshold of phase difference. To explain this setting, we consider two points denoted by two suffixes, $i$ and $j$. Considering the condition for detecting satellite points in \texttt{DBSCAN} is $(m_i - m_j)^2 + C^2 (\ell_i-\ell_j)^2 < \varepsilon^2$, the points $i$ and $j$ are clustered under our setting if they satisfy either $m_j=m_i$ and $|\ell_i-\ell_j| \leq 2\delta/\sqrt{3}$ or $m_j=m_i+1$ and $|\ell_i-\ell_j| \leq \delta$. The former case corresponds to a non-physical case where points with the same cycle are clustered. We can avoid this in most cases by setting the distance between the local minima with the same cycle to a large number in the distance matrix, which is an argument to the \texttt{DBSCAN} package. As a result, only the latter case occurs, and we can obtain the required clustering when we set $\delta = 0.10$. If a non-physical result such as over-clustering was obtained, the clustering process was repeated after reducing $\delta$ by 0.02 until reasonable results were achieved.

It should be noted that this analysis may incorrectly cluster distinct starspots that appear at the same phase. In such cases, $A(t)$ is considered to have multiple peaks. In this study, when a cluster contained a local minimum where $A(t_\mathrm{lm})$ was reduced by more than 40 percent from both of the two adjacent local maxima, we split the cluster before and after that local minimum and treated each as a distinct cluster.

\section{Estimation of Star-Spot Lifetime}
\label{app:scaling_tau}

Here, we introduce the procedure to estimate the lifetimes of starspots from sparse observation points, which are utilized for fitting. As in Appendix \ref{app:recurrent_sunspot}, we assume that sunspot evolution follows Equation \ref{eq:app_spot_evolution} with $\beta=0.3$. If $t_\mathrm{lm}$ and $A(t_\mathrm{lm})$ are obtained before and after $t_\mathrm{max}$, we can obtain $\tau_\mathrm{emerge}$ and $\tau_\mathrm{decay}$ as
\begin{align}
    \tau_\mathrm{emerge} = \frac{t_\mathrm{max} - t_\mathrm{lm}}{1-[A(t_\mathrm{lm})/A_\mathrm{max}]^{1-\beta}} \ \textrm{if} \ t_\mathrm{lm} \leq t_\mathrm{max} \label{eq:app_tau_emerge}
\end{align}
and
\begin{align}
    \tau_\mathrm{decay} = \frac{t_\mathrm{lm}-t_\mathrm{max}}{1-[A(t_\mathrm{lm})/A_\mathrm{max}]^{1-\beta}} \ \textrm{if} \ t_\mathrm{lm} \geq t_\mathrm{max}. \label{eq:app_tau_decay}
\end{align}
In our study, we calculated $\tau_\mathrm{emerge}$ and $\tau_\mathrm{decay}$ for each of the clustered local minima by substituting the local minimum with the smallest $A(t_\mathrm{lm})$ in the emergence phase into Equation (\ref{eq:app_tau_emerge}) and substituting the local minimum with the smallest $A(t_\mathrm{lm})$ in the decaying phase into Equation (\ref{eq:app_tau_decay}). As a result, we can obtain $\tau_\mathrm{life}$ using $\tau_\mathrm{life}=\tau_\mathrm{emerge}+\tau_\mathrm{decay}$. We note that the estimated $\tau_\mathrm{life}$ are subject to relative errors of up to 30 percent compared to the values with $\beta=0$ or $\beta=0.5$. However, this does not significantly affect the fit to Equation (\ref{eq:tau_scaling}) used in this study.

\section{Sun-as-a-star Analysis}
\label{app:method_sun-as-a-star}

\begin{figure}[t!]
\centering
\plotone{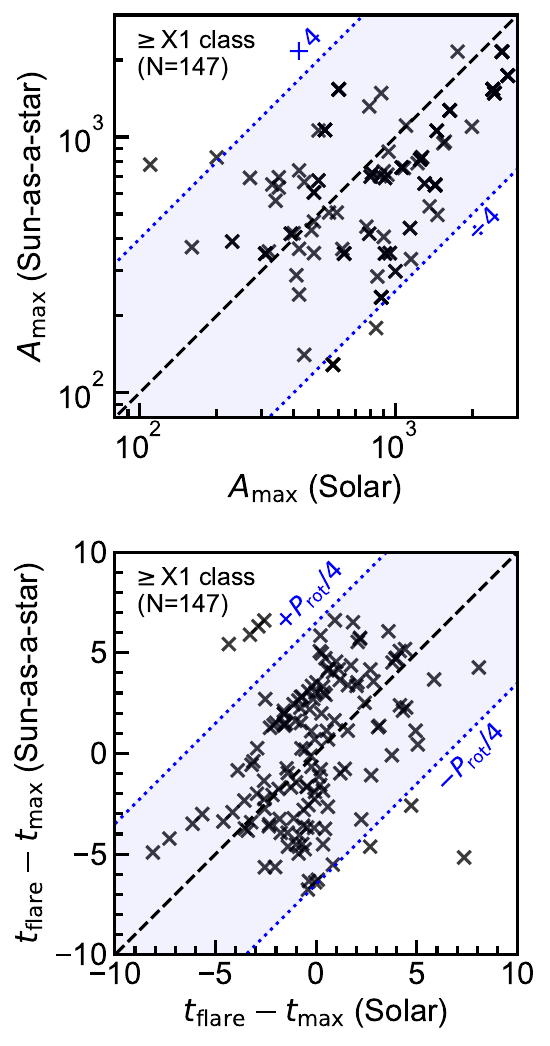}
\caption{(Upper) Direct comparison between the $A_\mathrm{max}$  of the solar analysis and $A_\mathrm{max}$ estimated by the sun-as-a-star analysis. The black dashed line marks 1:1 ratio. The blue areas separated by dotted lines indicate areas within a factor of 4. (Lower) The same figure as upper panel but for $t_\mathrm{max}-t_\mathrm{flare}$. The blue areas separated by dotted lines indicate areas within the deviation of $\pm P_\mathrm{rot}/4$. 
\label{fig:sun-as-a-star}}
\end{figure}

To evaluate the validity of the stellar analysis (Section \ref{subsec:method_stellar}), we conducted the same analysis using solar data \citep[see also][]{Namekata2019ApJ}. In this sun-as-a-star analysis, we use the total solar irradiance (TSI) obtained by the Variability of solar IRradiance and Gravity Oscillations (VIRGO) experiment\footnote{\url{ftp://ftp.pmodwrc.ch/pub/data/irradiance/virgo/TSI/}} \citep{Frohlich1995SoPh, Finsterle2021SciRep}, which is the spatially and spectrally integrated intensity of solar radiation at the Earth. To emulate the \textit{Kepler} light curve, we divide it into 90-day segments and normalize them by the median value for each segment, following \citet{Reinhold2020Sci}. For flare data, we use the X-class retrieved from the same flare catalog as for the solar analysis. To clarify the indeterminacy of the method, we analyze only the flares which have the same rotation cycle as the flare-productive sunspots reported by NOAA.

Figure \ref{fig:sun-as-a-star} shows the direct comparison between the values of solar analysis (see Section \ref{subsec:method_solar}) and the values estimated by the sun-as-a-star analysis for  $A_\mathrm{max}$ (upper panel) and $t_\mathrm{max}-t_\mathrm{flare}$ (lower panel). We found that the uncertainty of $A_\mathrm{max}$ induced by local-minima-tracing is mostly within the factor of 4, which is consistent with the previous analysis of \citet{Namekata2019ApJ}. Also, we found that the uncertainty of $A_\mathrm{max}$ induced by local-minima-tracing is mostly within $\pm P_\mathrm{rot}/4$. More specifically, more than 90 percent of the data in this sample fall within these ranges.

\bibliography{sample631}{}
\bibliographystyle{aasjournal}

\end{document}